\documentclass[twocolumn,preprintnumbers,superscriptaddress,nofootinbib,aps,prd,floatfix]{revtex4}

\usepackage{footmisc,enumerate}
\usepackage{subfigure}
\usepackage{amsmath,amssymb}
\usepackage{graphicx}
\usepackage{bbm,slashed}
\usepackage{xspace,slashed}
\usepackage{hyperref}
\hypersetup{colorlinks=true, citecolor=blue, urlcolor=blue, linkcolor=blue}
\usepackage[normalem]{ulem}

\begin{document}

\newcommand{\og}{\ensuremath{\tilde{O}_g}\xspace}
\newcommand{\ot}{\ensuremath{\tilde{O}_t}\xspace}

\def\ani#1{\textcolor{red}{#1}}

\providecommand{\abs}[1]{\lvert#1\rvert}

\newcommand{\Znunujets}{(Z\to{\nu\bar{\nu}})+\text{jets}}
\newcommand{\Welnujets}{(W\to{\ell\nu})+\text{jets}}
\newcommand{\Znunujet}{(Z\to{\nu\bar{\nu}})+\text{jet}}
\newcommand{\Welnujet}{(W\to{\ell\nu})+\text{jet}} 

\newcommand{\cw}{\ensuremath{C_{\widetilde{W}}\xspace}}
\newcommand{\chwb}{\ensuremath{C_{H\widetilde{W}B}}\xspace}

%%%%%%%%%%%%%%%%%%%%
%\title{Scrutinising the CP excesses}
\title{Extended Higgs sectors, effective field theory and Higgs phenomenology}
%%%%%%%%%%%%%%%%%%%%

\begin{abstract}
We consider the phenomenological implications of charged scalar extensions of the SM Higgs sector in addition to EFT couplings of this new state to SM matter. We perform a detailed investigation of modifications of loop-induced decays of the 125 GeV Higgs boson, which receives corrections from the propagating charged scalars alongside one-loop EFT operator insertions and demonstrate that the interplay of $H\to \gamma\gamma$ and $H\to Z\gamma$ decays can be used to clarify the additional states phenomenology in case a discovery is made in the future. In parallel, EFT interactions of the charged Higgs can lead to a decreased sensitivity to the virtual presence of charged Higgs states, which can significantly weaken the constraints that are naively expected from the precisely measured $H\to \gamma\gamma$ branching ratio. Again $H\to Z\gamma$ measurements provide complementary sensitivity that can be exploited in the future.
\end{abstract}

%%%%%%%%%%%%%%%%%%%%%%%%%%%%%%%%%%%%%%%%%%%%%%%%%%%%%%%%%%%%%%%%%
\author{Anisha} \email{anisha@iitk.ac.in}
\affiliation{Indian Institute of Technology Kanpur, Kalyanpur, Kanpur 208016, India\\[0.1cm]}
%%%%
\author{Upalaparna~Banerjee} \email{upalab@iitk.ac.in}
\affiliation{Indian Institute of Technology Kanpur, Kalyanpur, Kanpur 208016, India\\[0.1cm]}
%%%%
\author{Joydeep~Chakrabortty} \email{joydeep@iitk.ac.in}
\affiliation{Indian Institute of Technology Kanpur, Kalyanpur, Kanpur 208016, India\\[0.1cm]}
%%%%
\author{Christoph~Englert} \email{christoph.englert@glasgow.ac.uk}
\affiliation{School of Physics \& Astronomy, University of Glasgow, Glasgow G12 8QQ, United Kingdom\\[0.1cm]}
%%%%
\author{Michael~Spannowsky} \email{michael.spannowsky@durham.ac.uk}
\affiliation{Institute for Particle Physics Phenomenology, Durham University, Durham DH1 3LE, United Kingdom\\[0.1cm]}
%%%%%

%%%%%%%%%%%%%%%%%%%%%%%
\preprint{}
%%%%%%%%%%%%%%%%%%%%%%%
%%%%%%%%%%%%%%%%%%%%%%%

\maketitle
%%%%%%%%%%%%%%%%%%%%%%%%%%%%%%%%%%
\section{Introduction}
\label{sec:intro}
%%%%%%%%%%%%%%%%%%%%%%%%%%%%%%%%%%
The search for physics beyond the Standard Model (BSM) is the highest
priority of the Large Hadron Collider (LHC) after the discovery of the Higgs boson~\cite{Aad:2012tfa,Chatrchyan:2012ufa}.
So far, however, there has been no conclusive evidence for the presence
of new states or interactions. In parallel, the increasingly fine-grained
picture of the Higgs sector that ATLAS and CMS are obtaining creates a
phenomenological tension when Higgs data is contrasted with theoretically
well-motivated new physics models. In particular, Higgs sector extensions
that predict the presence of additional charged scalar bosons are constrained
by extra scalar contributions to the $H\to \gamma \gamma$ decay mode, 
which is experimentally clean, in addition to direct search constraints in the
context of, e.g., the two-Higgs-doublet or triplet models~\cite{Akeroyd:2012ms,Englert:2013zpa}.

In this work, we approach such constraints from a different perspective. We
extend the SM with an additional electromagnetically charged Higgs $h^\pm$ which acts as transparent
and minimal extension to the SM providing additional propagating degrees
of freedom to modify observed SM Higgs rates. These states appear in many BSM Higgs
sector extensions (e.g. in two-Higgs doublet~\cite{Branco:2011iw} or triplet models~\cite{Georgi:1985nv}). 
In parallel, we consider effective field theory (EFT) operators to parametrise interactions of the 
new charged state with the SM gauge sector model-independently, e.g., by integrating
out $h^\pm$-dominant interactions with and intrinsic EFT mass scale $\Lambda$.
Considering EFT operators up to dimension 6, we compute the loop-induced
decays of the SM Higgs~(see also~\cite{Hartmann:2015aia,Alonso:2013hga,Jenkins:2013wua,Jenkins:2013zja,Grojean:2013kd,Elias-Miro:2014eia,Elias-Miro:2013gya,Dawson:2018liq,Dawson:2018pyl}) in this scenario to identify regions of consistency with
the SM expectation (for similar analyses of electroweak observables see e.g.~\cite{Dawson:2018jlg,Dawson:2019clf}). This highlights the possibility of non-minimal interactions of the charged Higgs as parametrised by the EFT 
interactions to significantly reduce the sensitivity of the naively (highly) constraining SM Higgs
$H\to \gamma \gamma$ decay mode. In contrast, we will see that the phenomenologically challenging
$H\to Z\gamma$ branching can resolve cancellations that render the BSM $H\to 
\gamma\gamma$ decay consistent with the SM.

We organise this work as follows: In Sec.~\ref{sec:model}, we review the details
of the extended Higgs sector, introducing all relevant couplings and EFT operators 
that we consider in this work. Sec.~\ref{sec:calc} provides a short overview of our computation,
while we discuss constraints and results in Sec.~\ref{sec:results}. We conclude in Sec.~\ref{sec:conc}.

%%%%%%%%%%%%%%%%%%%%%%%%%%%%%%%%%%
\section{The model}\label{sec:model}
%%%%%%%%%%%%%%%%%%%%%%%%%%%%%%%%%%
For the purpose of our work we have considered the effective operators up to dimension 6, the full effective Lagrangian can be written as,  
\begin{eqnarray}\label{eq:1}
\mathcal{L} = \mathcal{L}_{\text{renorm}} +
\sum_{j=1}^{N}\frac{\mathcal{C}^{(5)}_j}{\Lambda}\mathcal{O}^{(5)}_j + \sum_{k=1}^{M}\frac{\mathcal{C}^{(6)}_k}{\Lambda^{2}}\mathcal{O}^{(6)}_k.
\end{eqnarray}
We extend the Higgs sector by considering an extra $SU(2)_L$ singlet scalar field $\mathcal{S}$ with hypercharge 1. The renormalisable part of the Lagrangian $\mathcal{L}_{\text{renorm}}$ mentioned in Eq.~\eqref{eq:1} then takes the form,
\begin{eqnarray}\label{eq:2}
\mathcal{L}_{\text{renorm}} &=& - \frac{1}{4}G^{A}_{\mu\nu}G^{A\mu\nu} - \frac{1}{4}W^{I}_{\mu\nu}W^{I\mu\nu} - \frac{1}{4}B_{\mu\nu}B^{\mu\nu}\nonumber\\ 
& &+(\mathcal{D}_{\mu} \phi)^{\dagger}(\mathcal{D}^{\mu} \phi)+ (\mathcal{D}_{\mu} \mathcal{S})^{\dagger}(\mathcal{D}^{\mu}\, \mathcal{S})-V(\phi,\mathcal{S})\nonumber\\
& &+i(\overline{L}\gamma^{\mu}\mathcal{D}_{\mu}L+\overline{e}\gamma^{\mu}\mathcal{D}_{\mu}e+\overline{Q}\gamma^{\mu}\mathcal{D}_{\mu}Q+\overline{u}\gamma^{\mu}\mathcal{D}_{\mu}u\nonumber\\
& &+\overline{d}\gamma^{\mu}\mathcal{D}_{\mu}d)
+ \mathcal{L}_{\text{Yukawa}}+h.c. ,
\end{eqnarray}
where $G^{A}_{\mu\nu},\,W^{I}_{\mu\nu}$ and $B_{\mu\nu}$ are the field strength tensors corresponding to $SU(3)_{C}, \,SU(2)_{L}$ and $U(1)_{Y}$ respectively. The generic form of the scalar potential $V(\phi,\mathcal{S})$ mentioned in Eq.~\eqref{eq:2},
\begin{eqnarray}\label{eq:3}
V(\phi,\mathcal{S}) &=& m_{1}^{2} (\phi^{\dagger}\phi)+m_{2}^{2} (\mathcal{S}^{\dagger}\mathcal{S})+\frac{\lambda_{1}}{2}(\phi^{\dagger}\phi)^{2}\nonumber\\
& &+\frac{\lambda_{2}}{2}(\mathcal{S}^{\dagger}\mathcal{S})^{2}+\frac{\lambda_{3}}{2}(\phi^{\dagger}\phi)(\mathcal{S}^{\dagger}\mathcal{S}).
\end{eqnarray}
The Yukawa part of the Lagrangian is also extended as the transformation properties of $\mathcal{S}$ under $SU(3)_C\otimes SU(2)_L\otimes U(1)_Y$ allow the interaction between the left-handed lepton doublets and the singlet scalar,
\begin{eqnarray}\label{eq:4}
\mathcal{L}_{\text{Yukawa}} &=& -y_{e}\overline{L}e\phi-y_{u}\overline{Q}u\tilde{\phi}-y_{d}\overline{Q}d\phi\nonumber\\
& &-f(\overline{L^{c}}i\tau_{2}L)\mathcal{S},
\end{eqnarray}
here, $\tilde{\phi}$ is the charge-conjugated SM Higgs doublet, %$\tilde{\phi}=i\,\tau_{2}\,\phi^{*}$
and $y_{e,u,d}$ are the Yukawa coupling matrices and $f$ is the coupling constant for the new interaction present in Eq.~\eqref{eq:4}.
In Eq.~\eqref{eq:1}, we also include effective operators that parametrise the new interactions between charged scalar and SM fields~\cite{Banerjee:2020jun}. 

In Tab.~\ref{table:mu>ea}, we have collected the operators that contribute to different rare and flavour-violating $l_{i} \to l_{j\neq i} \gamma$ processes. From the strong constraints on the decay width from these channels, we can infer that the Wilson coefficients corresponding to these operators are negligible, and we will not consider these operators in the remainder of this work.
%%%%%%%%%%%%%%%%%%%%%%%%%%%%%
\begin{table}[!b]
	\centering
	\renewcommand{\arraystretch}{1.9}
	{\scriptsize\begin{tabular}{||c|c||c|c||}
			\hline
			\hline
			\multicolumn{2}{||c||}{$\Psi^2\Phi^2$}&
			\multicolumn{2}{c||}{$\Psi^2\Phi^3$}\\
			\hline
			
			$\mathcal{O}_{ L e \phi \mathcal{S}} $&
			$\color{magenta}{\overline{L} \,e \,\tilde{\phi} \,\mathcal{S}}$&
			$\mathcal{O}_{L \phi \mathcal{S}} $&
			$\overline{L} \,e \,\phi \,(\mathcal{S}^{\dagger} \,\mathcal{S})$\\
			
			\hline
			\hline
			\multicolumn{2}{||c||}{$\Psi^2\Phi^2 \mathcal{D}$}&
			\multicolumn{2}{c||}{$\Psi^2\Phi X$}\\
			\hline
			
			$\mathcal{O}_{ \mathcal{S} L e \mathcal{D}} $&
			$(\overline{L^{c}} \,\gamma^{\mu} \,e) \,\tilde{\phi} \,(i\mathcal{D}_{\mu} \,\mathcal{S})$&
			$\mathcal{O}_{ e B \mathcal{S}} $&
			$B_{\mu\nu} (\,\overline{L^{c}} \,\sigma^{\mu\nu} \,L) \,\mathcal{S} $\\
			
			$\mathcal{O}_{\mathcal{S} L \mathcal{D}} $&
			$(\overline{L} \,\gamma^{\mu} \,L) \,(\mathcal{S}^{\dagger} \,i\overleftrightarrow{\mathcal{D}}_{\mu} \,\mathcal{S})$&
			$\mathcal{O}_{e W \mathcal{S}}$&
			$W^{I}_{\mu\nu} (\,\overline{L^{c}} \,\sigma^{\mu\nu} \,\tau^{I} \,L)  \,\mathcal{S} $\\
			
			$\mathcal{O}_{\mathcal{S} e \mathcal{D}} $&
			$(\overline{e} \,\gamma^{\mu} \,e) \,(\mathcal{S}^{\dagger} \,i\overleftrightarrow{\mathcal{D}}_{\mu} \,\mathcal{S})$&
			&\\
			\hline
	\end{tabular}}
	\caption{Effective operators relevant for $l_{i} \to l_{j\neq i} \gamma$ decay.  Operator written in magenta is the only dimension 5 operator that contributes to the decay amplitude. $\tau^I$ is $SU(2)$ generator and $I=1,2,3.$}
	\label{table:mu>ea}
\end{table}
The dimension 6 operators which contribute to $H\to \gamma \gamma$ and $H\to Z \gamma$ decay, have been tabulated in Tab.~\ref{table:h>aa&az} and the operators which contribute to $H\to gg$ decay are given in Tab.~\ref{table:h>gg}, in both cases no dimension 5 operator contributes.
%%%%%%%%%%%%%%%%%%%%%%%%%%%%%
\begin{table}[h]
	\centering
	\renewcommand{\arraystretch}{1.9}
	{\scriptsize\begin{tabular}{||c|c||c|c||}
			\hline
			\hline
			\multicolumn{2}{||c||}{$\Phi^4\mathcal{D}^2$}&
			\multicolumn{2}{c||}{$\Phi^6$}\\
			\hline
			
			$\mathcal{O}_{ \mathcal{S} \phi\mathcal{D}} $&
			$(\mathcal{S}^{\dagger}\,\mathcal{S})\,\left[(\mathcal{D}^{\mu}\,\phi)^{\dagger}(\mathcal{D}_{\mu}\,\phi)\right]$&
			$\mathcal{O}_{\phi\mathcal{S}} $&
			$(\phi^{\dagger} \,\phi)^2 \,(\mathcal{S}^{\dagger} \,\mathcal{S})$\\
			
			$\mathcal{O}_{\phi \mathcal{S} \mathcal{D}}$&
			$(\phi^{\dagger}\,\phi)\,\left[(\mathcal{D}^{\mu}\,\mathcal{S})^{\dagger}(\mathcal{D}_{\mu}\,\mathcal{S})\right]$&
			&
			\\
			\hline
			\hline
			\multicolumn{4}{||c||}{$\Phi^2X^2$}\\
			\hline
			
			$\mathcal{O}_{ B \phi} $&
			$\color{blue}{B_{\mu\nu} \,B^{\mu\nu} \,(\phi^{\dagger} \,\phi)}$&
			$\mathcal{O}_{ B \mathcal{S}} $&
			$B_{\mu\nu} \,B^{\mu\nu} \,(\mathcal{S}^{\dagger} \,\mathcal{S})$\\
			
			$\mathcal{O}_{\widetilde{B} \phi} $&
			$\color{blue}{\widetilde{B}_{\mu\nu} \,B^{\mu\nu} \,(\phi^{\dagger} \,\phi)}$&
			$\mathcal{O}_{\widetilde{B} \mathcal{S}} $&
			$\widetilde{B}_{\mu\nu} \,B^{\mu\nu} \,(\mathcal{S}^{\dagger} \,\mathcal{S})$\\
			
			$\mathcal{O}_{ W \phi} $&
			$\color{blue}{W^{I}_{\mu\nu} \,W^{I\mu\nu} \,(\phi^{\dagger} \,\phi)}$&
			$\mathcal{O}_{ W \mathcal{S}} $&
			$W^{I}_{\mu\nu} \,W^{I\mu\nu} \,(\mathcal{S}^{\dagger} \,\mathcal{S})$\\

			$\mathcal{O}_{ \widetilde{W} \phi } $&
			$\color{blue}{\widetilde{W}^{I}_{\mu\nu} \,W^{I\mu\nu} \,(\phi^{\dagger} \,\phi)}$&
			$\mathcal{O}_{ \widetilde{W} \mathcal{S} } $&
			$\widetilde{W}^{I}_{\mu\nu} \,W^{I\mu\nu} \,(\mathcal{S}^{\dagger} \,\mathcal{S})$
			\\

			$\mathcal{O}_{ W B \phi } $&
			$\color{blue}{W^{I}_{\mu\nu} \,B^{\mu\nu} \,(\phi^{\dagger} \,\tau^{I}\,\phi)}$&
			& 
			\\
			
			$\mathcal{O}_{ \widetilde{W} B \phi } $&
			$\color{blue}{\widetilde{W}^{I}_{\mu\nu} \,B^{\mu\nu} \,(\phi^{\dagger} \,\tau^{I}\,\phi)}$&
			&
			\\
			\hline
	\end{tabular}}
	\caption{Effective operators relevant for $H\to \gamma \gamma$ and $H\to Z \gamma$ decay.  Operators in blue are the  pure SMEFT operators that contribute to the LO amplitudes in $H$ decay. $\tau^I$ is $SU(2)$ generator and $I=1,2,3.$ The dual field strengths are defined as $\widetilde{X}_{\mu\nu}=\epsilon_{\mu\nu\rho\delta} X^{\rho\delta}/2$.}
	\label{table:h>aa&az}
\end{table}

\begin{table}[h]
	\centering
	\renewcommand{\arraystretch}{1.9}
	{\scriptsize\begin{tabular}{||c|c||c|c||}
			\hline
			\hline
			\multicolumn{2}{||c||}{$\Phi^4\mathcal{D}^2$}&
			\multicolumn{2}{c||}{$\Phi^6$}\\
			\hline
			
			$\mathcal{O}_{\phi \mathcal{S} \mathcal{D}}$&
			$(\phi^{\dagger}\,\phi)\,\left[(\mathcal{D}^{\mu}\,\mathcal{S})^{\dagger}(\mathcal{D}_{\mu}\,\mathcal{S})\right]$&
			$\mathcal{O}_{\phi\mathcal{S}} $&
			$(\phi^{\dagger} \,\phi)^2 \,(\mathcal{S}^{\dagger} \,\mathcal{S})$\\
			\hline
			\hline
			\multicolumn{4}{||c||}{$\Phi^2X^2$}\\
			\hline
			
			$\mathcal{O}_{G \phi} $&
			$\color{blue}{G^{A}_{\mu\nu}\,G^{A\mu\nu}\,(\phi^{\dagger} \,\phi)}$&
			$\mathcal{O}_{G \mathcal{S}} $&
			$G^{A}_{\mu\nu}\,G^{A\mu\nu}\,(\mathcal{S}^{\dagger} \,\mathcal{S})$\\
			
			$\mathcal{O}_{\widetilde{G} \phi } $&
			$\color{blue}{\widetilde{G}^{A}_{\mu\nu}\,G^{A\mu\nu}\,(\phi^{\dagger} \,\phi)}$&
			$\mathcal{O}_{\widetilde{G} \mathcal{S} } $&
			$\widetilde{G}^{A}_{\mu\nu}\,G^{A\mu\nu}\,(\mathcal{S}^{\dagger} \,\mathcal{S})$\\
			
			\hline
	\end{tabular}}
	\caption{Effective operators relevant for $H\to g g$ decay.  Operators in blue are the  pure SMEFT operators that contribute to the LO amplitudes in $H$ decay, $A=1,2...8$. The dual field strength tensors are in the convention of Tab.~\ref{table:h>aa&az}.}
	\label{table:h>gg}
\end{table}
%%%%%%%%%%%%%%%%%%%%%%%%%%%%%

After spontaneous symmetry breaking (SSB), the $SU(2)_L$ doublet scalar $\phi$ takes the following form and produces the physical neutral Higgs $H$, 
\begin{eqnarray}\label{eq:5}
\phi =
\begin{bmatrix}
G^{+}\\
\frac{1}{\sqrt{2}}(v+H+i\,G^{0})	
\end{bmatrix},
\end{eqnarray}
where $G^{\pm}$ and $G^{0}$ are the charged and neutral Goldstone fields respectively. After SSB, the $SU(2)_L$ singlet scalar $\mathcal{S} \,(\mathcal{S}^{\dagger})$ emerges as charged scalar field $h^{\pm}$. 
The operator $(\phi^{\dagger}\,\phi)\,\left[(\mathcal{D}^{\mu}\,\mathcal{S})^{\dagger}(\mathcal{D}_{\mu}\,\mathcal{S})\right]$ changes the normalisation of the $h^{\pm}$-kinetic term. We can redefine the field as,
\begin{eqnarray}\label{eq:6}
h^{\pm} \to (1-\mathcal{C}_{\phi\mathcal{S}\mathcal{D}}\frac{v^2}{2\Lambda^2})\,h^{\pm},
\end{eqnarray}
to recover a canonically normalised Lagrangian. The mass of charged scalar $h^{\pm}$ receives contributions from the effective operator $(\phi^{\dagger} \,\phi)^2 \,(\mathcal{S}^{\dagger} \,\mathcal{S})$. Considering this operator and the proper redefinition of field given in Eq.~\eqref{eq:6}, we find the squared of the mass of $h^{\pm}$ from Eq.~\eqref{eq:3} to be
\begin{eqnarray}\label{eq:7}
M_{h^{\pm}}^{2} &=& (m_2^2+\frac{\lambda_{3}}{2}v^2) \,(1-\mathcal{C}_{\phi\mathcal{S}\mathcal{D}}\frac{v^2}{\Lambda^2})\,+\,\mathcal{C}_{\phi\mathcal{S}}\frac{v^4}{\Lambda^2}\,.
\end{eqnarray}

The $\Phi^6$ class of operators (in the convention of~\cite{Grzadkowski:2010es}) are given by $(\mathcal{S}^{\dagger}\mathcal{S})^3$, $(\phi^{\dagger} \,\phi)^2 \,(\mathcal{S}^{\dagger} \,\mathcal{S})$ and $(\phi^{\dagger} \,\phi) \,(\mathcal{S}^{\dagger} \,\mathcal{S})^2$ and these parametrise the new interactions of the new scalar and the SM Higgs. None of these operators impact the stability of the neutral vacuum.

%%%%%%%%%%%%%%%%%%%%%%%%%%%%%%%%%%
\begin{figure*}
\includegraphics[width=0.9\textwidth]{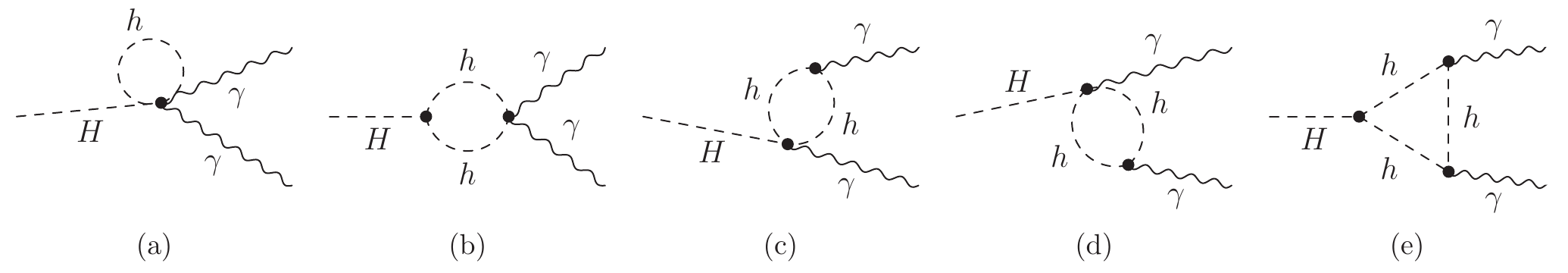}
\caption{BSM Feynman topologies contributing to the Higgs di-photon branching $H\to \gamma\gamma$ via the new propagating $h^\pm$ and its EFT interactions, e.g. the 5-point interaction of (a) but also modifications of 3- and 4-point interactions. \label{fig:feyn}}
\end{figure*}
%%%%%%%%%%%%%%%%%%%%%%%%%%%%%%%%%%

%%%%%%%%%%%%%%%%%%%%%%%%%%%%%%%%%%
\section{Elements of the Calculation}\label{sec:calc}
%%%%%%%%%%%%%%%%%%%%%%%%%%%%%%%%%%
As detailed in the previous section, we start with a canonically normalised effective Lagrangian in the broken electroweak phase. The calculation of the loop-induced decays of SM Higgs decay then receives contributions from the propagating SM degrees of freedom, the BSM charged scalar $h^\pm$, as well as the dimension 6 operator insertions at one loop. The latter radiatively induce the SMEFT operators tabulated under $\Phi^2 X^2$ class in the first column of Tab.~\ref{table:h>aa&az},
	\begin{multline}
		\label{eq:bare_unphysical}
		{\cal{L}}_{\text{d6}} = \\
		{v H \over \Lambda^2} \left( {\mathcal{C}}_{W \phi } W^{I\mu\nu} W^I_{\mu\nu} - {\mathcal{C}}_{WB\phi} W^{3\mu\nu} B_{\mu\nu} +{\mathcal{C}}_{B\phi}  B^{\mu\nu} B_{\mu\nu}  \right)\\
		+ {v H \over \Lambda^2}  \left( {\mathcal{C}}_{\widetilde W\phi} W^{3\mu\nu} \widetilde W^3_{\mu\nu} - {\mathcal{C}}_{\widetilde W B\phi} B^{\mu\nu} \widetilde W^3_{\mu\nu}  + {\mathcal{C}}_{\widetilde B\phi} B^{\mu\nu} \widetilde B_{\mu\nu}  \right)\\
		+ {v H \over \Lambda^2} \left( {\mathcal{C}}_{G\phi} G^{a,\mu\nu} G^a_{\mu\nu}  + {\mathcal{C}}_{\widetilde G\phi} G^{a,\mu\nu} \widetilde{G}^a_{\mu\nu}  \right),
	\end{multline}
	which need to be included as part of the renormalisation of the processes that we consider in this work. 	
	The above ${\cal{L}}_{\text{d6}}$ is given in terms of mass eigenstates using  
	\begin{equation}
		\left(
		\begin{matrix}
			W^3_{\mu}\\
			B_{\mu}
		\end{matrix}
		\right) 
		=
		%\rightarrow
		\left(
		\begin{matrix}
			\frac{g_{_{W}}}{\sqrt{g^2_{_{W}} + g^2_{_{Y}}}} & 	\frac{g_{_{Y}}}{\sqrt{g^2_{_{W}}+ g^2_{_{Y}}}} \\
			-\frac{g_{_{Y}}}{\sqrt{g^2_{_{W}} + g^2_{_{Y}}}}  & 	\frac{g_{_{W}}}{\sqrt{g^2_{_{W}} + g^2_{_{Y}}}} \\ 
		\end{matrix}
		\right)
		\left(
		\begin{matrix}
			Z_{\mu}\\
			A_{\mu}
		\end{matrix}
		\right)\,,
	\end{equation}
	where, $g_{_{W}}$ and $g_{_{Y}} $ are the gauge coupling constants corresponding to the $SU(2)_{L}$ and $U(1)_{Y}$ respectively, as
	\begin{multline}
		\label{eq:bare}
		{\cal{L}}_{\text{d6}} = {v H \over \Lambda^2} \left( {\mathcal{C}}_{A\phi } A^{\mu\nu} A_{\mu\nu} + {\mathcal{C}}_{AZ\phi} A^{\mu\nu} Z_{\mu\nu}  +  {\mathcal{C}}_{Z\phi}  Z^{\mu\nu} Z_{\mu\nu}  \right)\\
		+ {v H \over \Lambda^2}  \left( {\mathcal{C}}_{\widetilde A\phi} A^{\mu\nu} \widetilde A_{\mu\nu} + {\mathcal{C}}_{A\widetilde Z\phi} A^{\mu\nu} \widetilde Z_{\mu\nu}  + {\mathcal{C}}_{\widetilde Z\phi} Z^{\mu\nu} \widetilde Z_{\mu\nu}  \right)\\
		+ {v H \over \Lambda^2} \left( {\mathcal{C}}_{G\phi} G^{a,\mu\nu} G^a_{\mu\nu}  + {\mathcal{C}}_{\widetilde G\phi} G^{a,\mu\nu} \widetilde{G}^a_{\mu\nu}  \right),
	\end{multline}
	and 
	\begin{equation}
\begin{split}
{\mathcal{C}}_{A\phi }& =\frac{g_{_{Y}}^2 }{g^2_{_{W}} + g^2_{_{Y}}}  {\mathcal{C}}_{W\phi }  -\frac{g_{_{W}} g_{_{Y}}}{g^2_{_{W}} + g^2_{_{Y}}} {\mathcal{C}}_{WB\phi } + \frac{g_{_{W}}^2}{g^2_{_{W}} + g^2_{_{Y}}} {\mathcal{C}}_{B\phi }, \nonumber \\
{\mathcal{C}}_{AZ\phi }& =\frac{2 g_{_{W}} g_{_{Y}}}{g^2_{_{W}} + g^2_{_{Y}}}  ({\mathcal{C}}_{W\phi }-{\mathcal{C}}_{B\phi }) -\frac{g^2_{_{W}} - g^2_{_{Y}}}{g^2_{_{W}} + g^2_{_{Y}}} {\mathcal{C}}_{WB\phi }, \nonumber \\
{\mathcal{C}}_{Z\phi }& =\frac{g_{_{W}}^2 }{g^2_{_{W}} + g^2_{_{Y}}}  {\mathcal{C}}_{W\phi }  +\frac{g_{_{W}} g_{_{Y}}}{g^2_{_{W}} + g^2_{_{Y}}} {\mathcal{C}}_{WB\phi } + \frac{g_{_{Y}}^2}{g^2_{_{W}} + g^2_{_{Y}}} {\mathcal{C}}_{B\phi }. \nonumber
\end{split}
\end{equation}
Similar relations hold for the CP-odd operators. 

In the following we will sketch the calculation of the $H\to \gamma \gamma$ branching. The $H\to Z\gamma, gg$ decay results can be obtained in a similar fashion, but we will comment on process-specific differences where they are relevant. In the case of the  $H(k_1)\to \gamma (k_2) \gamma (k_3)$ amplitude this gives rise to a new tree-level contribution
\begin{equation}
\parbox{2.5cm}{\includegraphics[width=2.5cm]{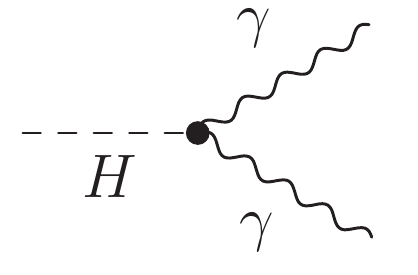}}
\to {\cal{M}}_{\text{LO}}={ {\mathcal{C}}_{\widetilde A\phi}  v\over \Lambda^2} O_o + { {\mathcal{C}}_{A\phi}  v\over \Lambda^2} O_e,
\end{equation}
with
\begin{align}
 O_o  &= -4 \epsilon^{\mu\nu\rho\delta} \epsilon^\ast_\mu(k_2)\epsilon^\ast_\nu(k_3) k_{1,\rho} k_{2,\delta}\,, \\
 O_e  &= 2 \left[2(\epsilon^\ast(k_2) k_1) (\epsilon^\ast(k_3) k_1) - k_1^2 (\epsilon^\ast(k_2)\epsilon^\ast(k_3) ) \right]\,,
\end{align}
and it is these operator structures that will be renormalised as a consequence of the one-loop $h^\pm$-related EFT insertions, while the renormalisable interactions of the propagating $h^\pm$  lead to a ultraviolet (UV)-finite modification of the $H\to \gamma \gamma$ partial decay width.

Throughout this work we chose on-shell renormalisation for SM fields alongside the $\overline{\text{MS}}$ scheme of Wilson coefficients. The Lagrangian of Eq.~\eqref{eq:bare} leads to a counter term contribution for the $H\to \gamma \gamma$ amplitude
\begin{multline}
\label{eq:cts}
{\cal{M}}_{\text{CT}} =\\
\left[ \delta {\mathcal{C}}_{A\phi} + {\mathcal{C}}_{A\phi} \left( \delta Z_H/2 + \delta Z_{AA}\right) +  {\mathcal{C}}_{AZ\phi} \delta Z_{ZA}/2   \right]{v\,O_e\over \Lambda^2}  \\
+\left[ \delta  {\mathcal{C}}_{\widetilde A\phi} +{\mathcal{C}}_{\widetilde A\phi} \left( \delta Z_H/2 + \delta Z_{AA} \right) +  {\mathcal{C}}_{A\widetilde Z\phi} \delta Z_{ZA}/2  \right] {v\,O_o\over \Lambda^2}\\
+ {\delta v \over \Lambda^2 } {\cal{M}}_{\text{LO}} \,,
\end{multline}
where the implications of $Z-\gamma$ mixing have been included. These $\delta Z_{H}$ and $\delta Z_{VV'}$ factors correspond to the renormalisation constants of the $H\to H$ and $V\to V'$ two-point functions respectively and result from a replacement of the bare quantities
\begin{equation}
\left(
\begin{matrix}
Z\\
A
\end{matrix}
\right) 
\rightarrow
\left(
\begin{matrix}
 1 +{1\over 2} \delta Z_{ZZ} & {1\over 2} \delta Z_{ZA} \\
 {1\over 2} \delta Z_{AZ}  & 1 +{1\over 2} \delta Z_{AA} \\ 
\end{matrix}
\right)
\left(
\begin{matrix}
Z\\
A
\end{matrix}
\right)\,.
\end{equation}
They are determined by imposing on-shell conditions on the real parts of the gauge boson self-energies, see e.g.~\cite{Denner:1991kt}. The dimension 6 counter terms  $\delta {\mathcal{C}}$  arise from shifting the bare Wilson coefficients in Eq.~\eqref{eq:bare}:  $ {\mathcal{C}} \to  {\mathcal{C}}  + \delta {\mathcal{C}}$ while $\delta v = - \delta T/M_H^2$ is related to the tadpole counter term~\cite{Fleischer:1980ub,Denner:2019vbn}. The explicit expressions of these counter terms have been given in the Appendix~\ref{sec:appendix} for the case of $H\to \gamma \gamma$.

As indicated in these equations, we include loop corrections and renormalisation constants evaluated to order $\Lambda^{-2}$, i.e. we strictly work in the dimension 6 framework such that the considered field theory is {\emph{technically}} renormalisable. We modified the {\textsc{SmeftFR}} package~\cite{Dedes:2019uzs} to add the charged scalar in our calculation and included all relevant Feynman diagrams from a {\sc{FeynRules}}~\cite{Alloul:2013bka}-generated model file using a modified version {\sc{FeynArts}}~\cite{Hahn:2000kx} to include interactions up to 6-point vertices. This is essential for a consistent result as the diagram of Fig.~\ref{fig:feyn} (a) is related to other EFT-$h^\pm$ diagrams by gauge symmetry. 

While only the BSM contributions to $H\to \gamma \gamma$ are sketched in Fig.~\ref{fig:feyn}, we include the SM contributions throughout, in particular for cross-checks against SM results at analytical~\cite{Hahn:2000kx} (we use Feynman gauge throughout our calculation) as well as numerical level by comparing to the results reported by the Higgs cross-section working group~\cite{Dittmaier:2011ti,Dittmaier:2012vm,Heinemeyer:2013tqa,deFlorian:2016spz}. Identical cross-checks were performed for the $H\to Z\gamma$ and $H\to gg$ decay calculations\footnote{In the $H\to Z\gamma$ case the SM part of the amplitude receives an additional counter term contribution $\sim e m_W/(2 s_\theta c_\theta) \delta Z_{AZ}$ from $Z-\gamma$ mixing, where $s_\theta,c_\theta$ are the sine and cosine of the Weinberg angle, respectively.}. 

%%%%%%%%%%%%%%%%%%%%%%%%%%%%%%%%%%
\begin{figure*}[!p]
\subfigure[\label{fig:dec1a}]{\includegraphics[width=0.49\textwidth]{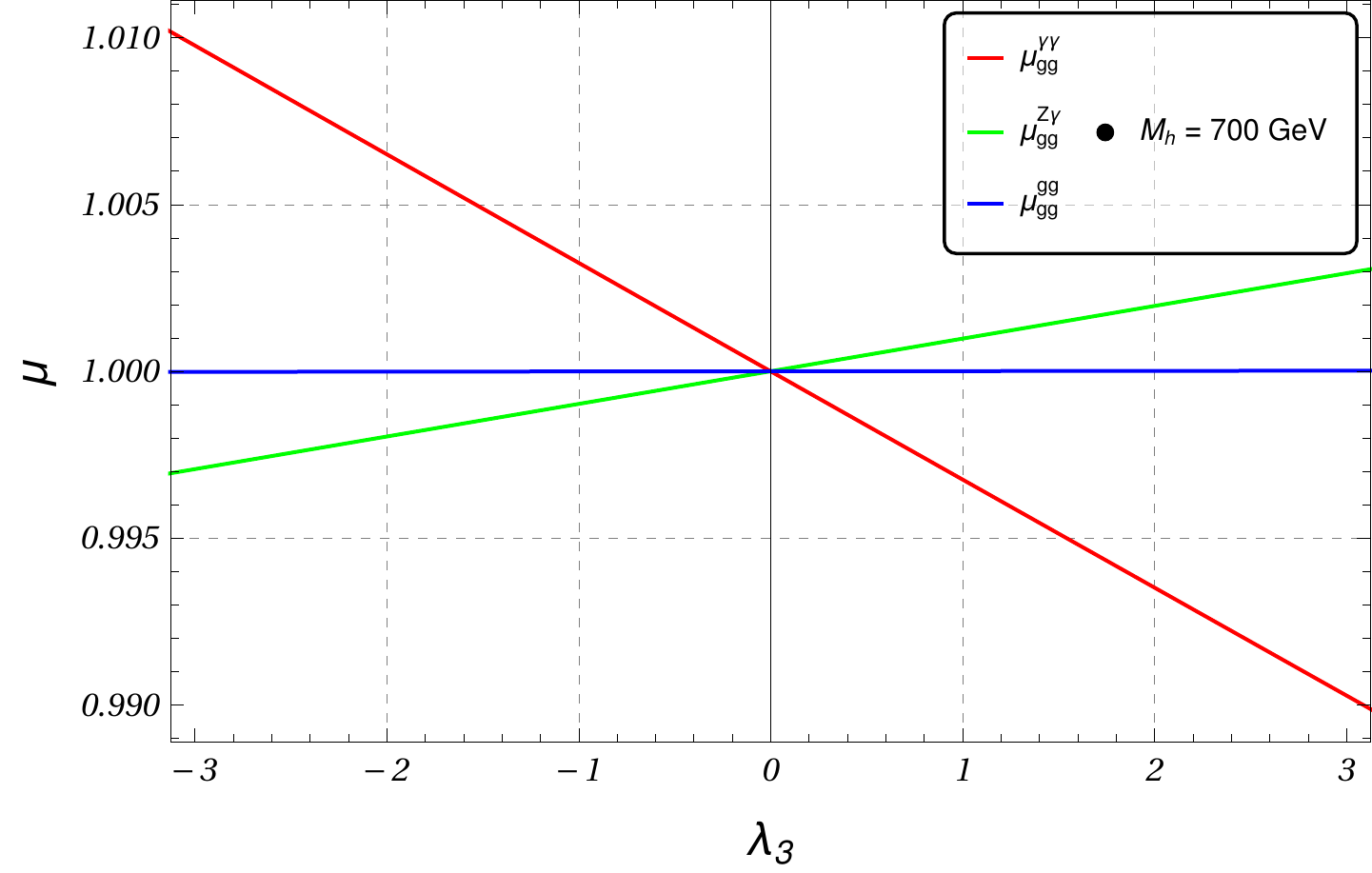}}\hfill
\subfigure[\label{fig:dec2b}]{\includegraphics[width=0.49\textwidth]{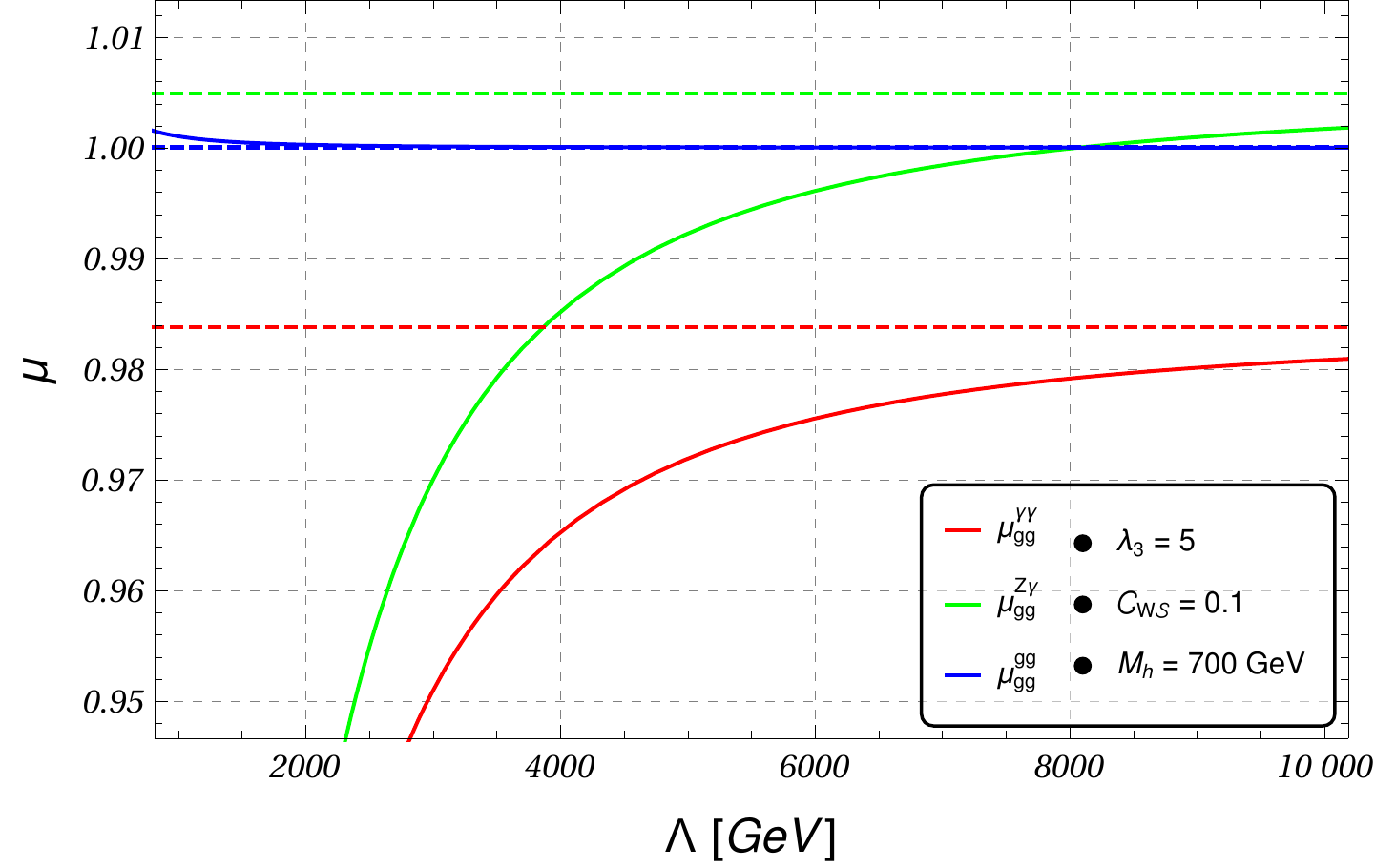}}
\caption{Approaching the SM as a function of $\lambda_3$ (a) for zero Wilson coefficient and (b) for non-trivial Wilson coefficient choices. The dashed lines represent modifications to the decay widths from a charged Higgs without higher-dimensional interactions, which is asymptotically approached for $\Lambda\to \infty$ for non-vanishing Wilson coefficient choices. \label{fig:dec1}}
\end{figure*}

As mentioned in the introduction, in this work, we will assume that new physics is dominantly related to the $h^\pm$ bosons' interactions, i.e., all SMEFT operators will be sourced radiatively through $h^\pm$ operators; the UV-singular structure of Fig.~\ref{fig:feyn} is only related to the operator matrix elements $O_o,O_e$. Furthermore, only the CP-odd (even) operators of Tab. \ref{table:h>aa&az} contribute to UV singularities dressing the $O_o$ ($O_e$) amplitudes at one-loop level. Adding the counter-term contributions to all one-loop diagrams of Fig.~\ref{fig:feyn}, we can therefore consistently absorb all singularities of the BSM one-loop correction into a redefinition of the SMEFT operators in the mass basis shown in Eq.~\eqref{eq:bare}.

The amplitude contains scalar two- and three-point functions $B_0,C_0$ in the convention of Passarino and Veltman~\cite{Passarino:1978jh,Denner:2005nn} which we include analytically in the case of $H\to \gamma \gamma,gg$, where we deal with a two-scale problem. In the case of $H\to Z\gamma$ we evaluate the three scale $C_0$ function numerically using {\sc{LoopTools}}~\cite{Hahn:1998yk,Hahn:2000jm}. As done in the SM, we include the full squared amplitude of the (renormalised) one-loop result to the calculation of the respective decay widths. We will see, however, that for perturbative parameter choices, the dependence of physical results is well-approximated by linearised Wilson coefficient dependencies. The phase space integration is straightforward and can be performed analytically~\cite{Zyla:2020zbs}. 

\medskip 

We are particularly interested in the modifications of the loop-induced $\gamma\gamma,Z\gamma, gg$ decays to the total Higgs decay width and the resulting branching ratio modifications, as well as SM Higgs production via the dominant gluon fusion (GF) channel. To this end we choose vanishing values for the renormalised Wilson coefficients of Eq.~\eqref{eq:bare}, which otherwise would impact the Higgs phenomenology already at tree-level. The leading order approximation of the GF cross section scales as (see e.g. \cite{Djouadi:2005gi})
\begin{equation}
{\sigma^\text{BSM}_{\text{GF}} \over {\sigma^\text{SM}_{\text{GF}}}} = {\Gamma^{\text{BSM}}(H\to gg) \over \Gamma^{\text{SM}}(H\to gg) },
\end{equation}
where $\Gamma^i(H\to gg)$ represents the different partial decay widths of $H\to gg$. Branching ratios are modified as
\begin{equation}
{ {\text{BR}}^{\text{BSM}}(H\to X) \over {\text{BR}}^{\text{SM}}(H\to X) }
= {  {\Gamma^{\text{BSM}}(H\to X) \over {\Gamma^{\text{SM}}(H\to X) } }}
{  { \Gamma^{\text{SM}}_{\text{tot}}} \over  { \Gamma^{\text{BSM}}_{\text{tot}}}},
\end{equation}
where the total decay widths are
\begin{equation}
\Gamma^{\text{(B)SM}} = \sum_X \Gamma^{\text{(B)SM}}(H\to X)\,.
\end{equation}
Assuming the narrow width approximation, the 125 GeV Higgs signal strength is then given by
\begin{equation}
\mu^X_{gg} = {[{\sigma_{\text{GF}} \times \text{BR}(H\to X) ]^\text{BSM}} \over
[{\sigma_{\text{GF}} \times \text{BR}(H\to X) ]^\text{SM}} }\,.
\end{equation}

\medskip 

It is interesting to see how the SM result is obtained as a function of the new degrees of freedom and the higher-dimensional operator contributions. Firstly, for all ${\cal{C}}_i=0$, the new physics contributions are controlled by $\lambda_3$ alone and as can be seen in Fig.~\ref{fig:dec1a}, we obtain the SM expectation irrespective of $M_h$ for $\lambda_3$. We can also see that for perturbative coupling choices, we are dominated by a linear behaviour of the new physics coupling. Secondly, the Appelquist-Carazzone decoupling theorem~\cite{Appelquist:1974tg} implies an asymptotic SM result for $M_h\gg M_H$. While these results are known from concrete models with propagating degrees of freedom such as, e.g., the two-Higgs-doublet model~\cite{Gunion:1989we, Gunion:2002zf}, the contribution of the EFT operators is shown in Fig.~\ref{fig:dec2b}. For $\lambda_3=0$ we can directly observe the decoupling of new physics when the cut-off scale is removed from the theory $\Lambda \to \infty$. For $\lambda_3\neq 0$, we asymptotically approach the results that include the propagating $h^\pm$, Fig.~\ref{fig:dec2b} highlighted by the dashed lines.

%%%%%%%%%%%%%%%%%%%%%%%%%%%%%%%%%%
\section{Results}\label{sec:results}
%%%%%%%%%%%%%%%%%%%%%%%%%%%%%%%%%%

%%%%%%%%%%%%%%%%%%%%%%%%%%%%%%%%%%
\begin{figure*}[!p]
	\subfigure[]{\includegraphics[width=0.45\textwidth]{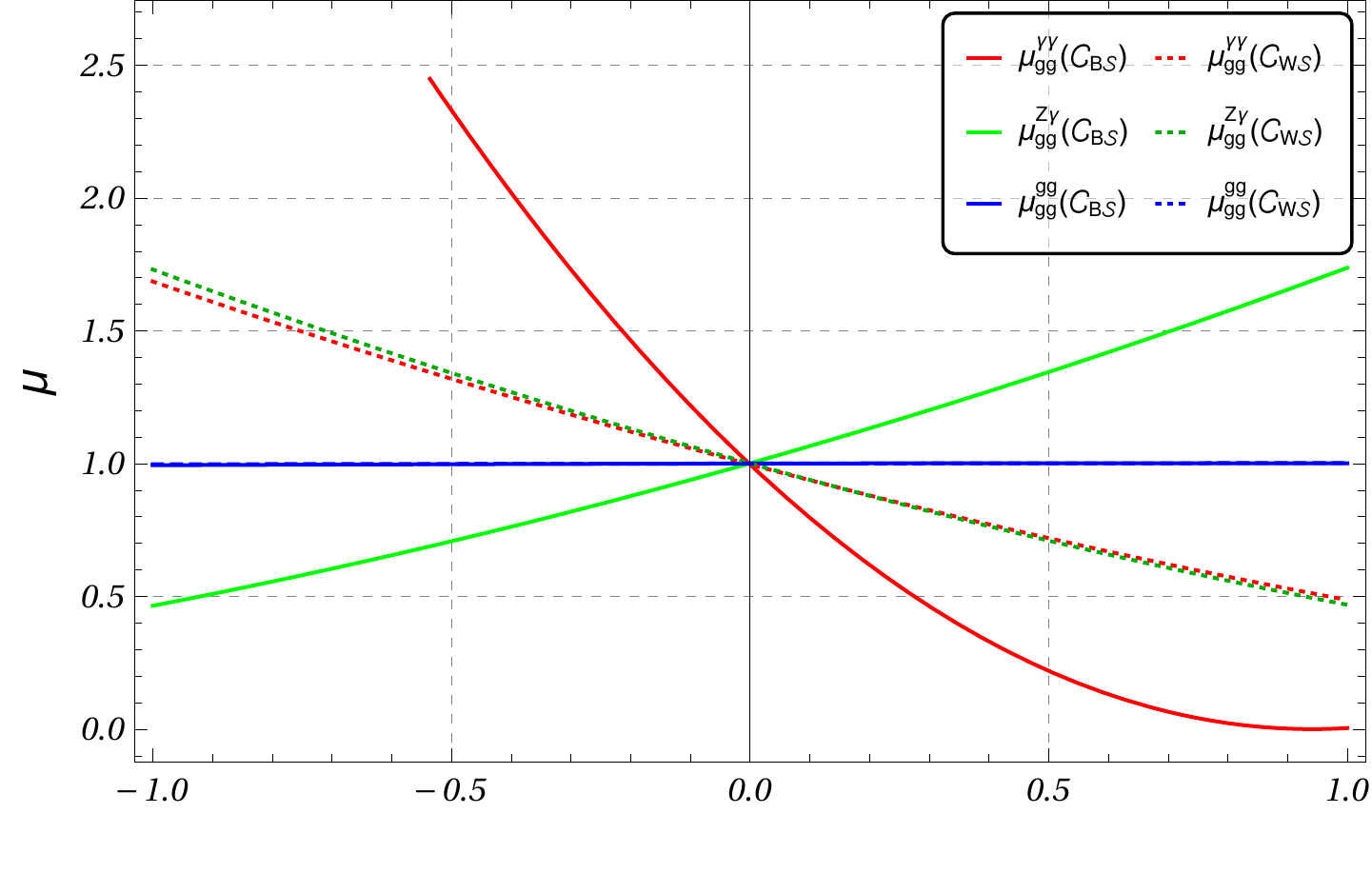}}\hfill %${\cal{C}}_{BS}$
	\subfigure[]{\includegraphics[width=0.45\textwidth]{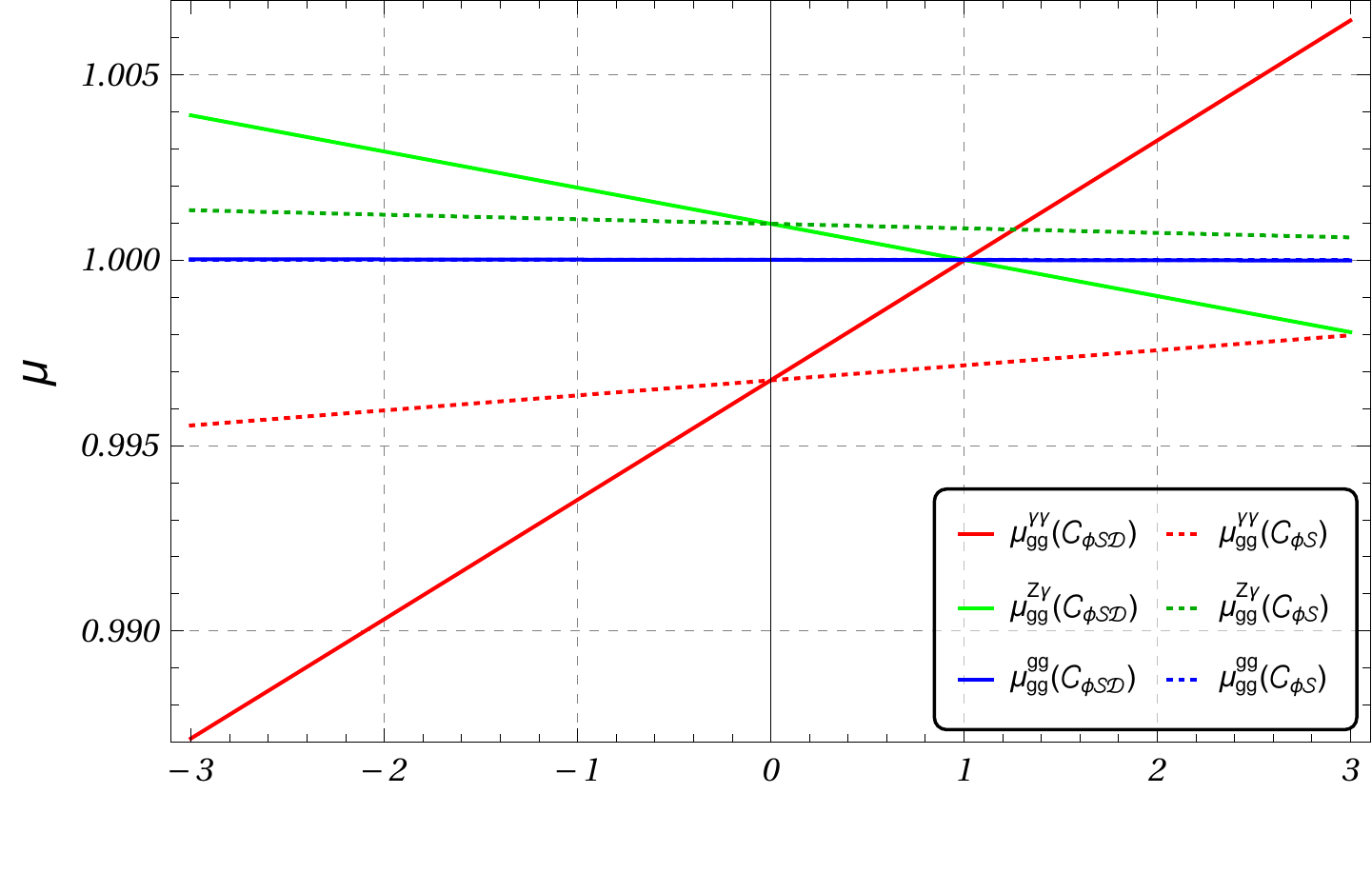}}\\ %$C_{\Phi S},C_{\Phi SD}$
	\subfigure[]{\includegraphics[width=0.45\textwidth]{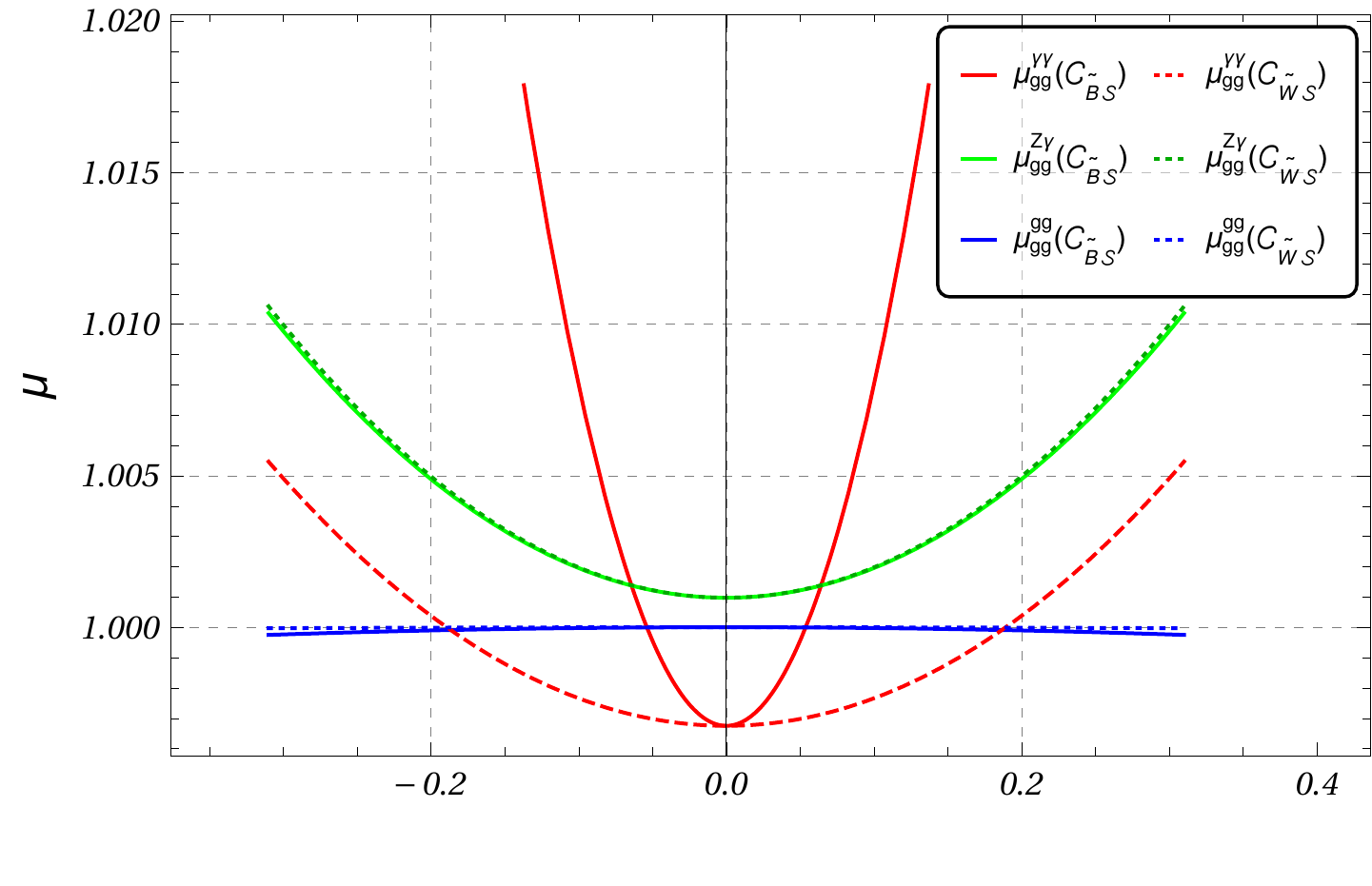}}\hfill %$\widetilde C_{WS}, \widetilde C_{BS}$
	\subfigure[]{\includegraphics[width=0.45\textwidth]{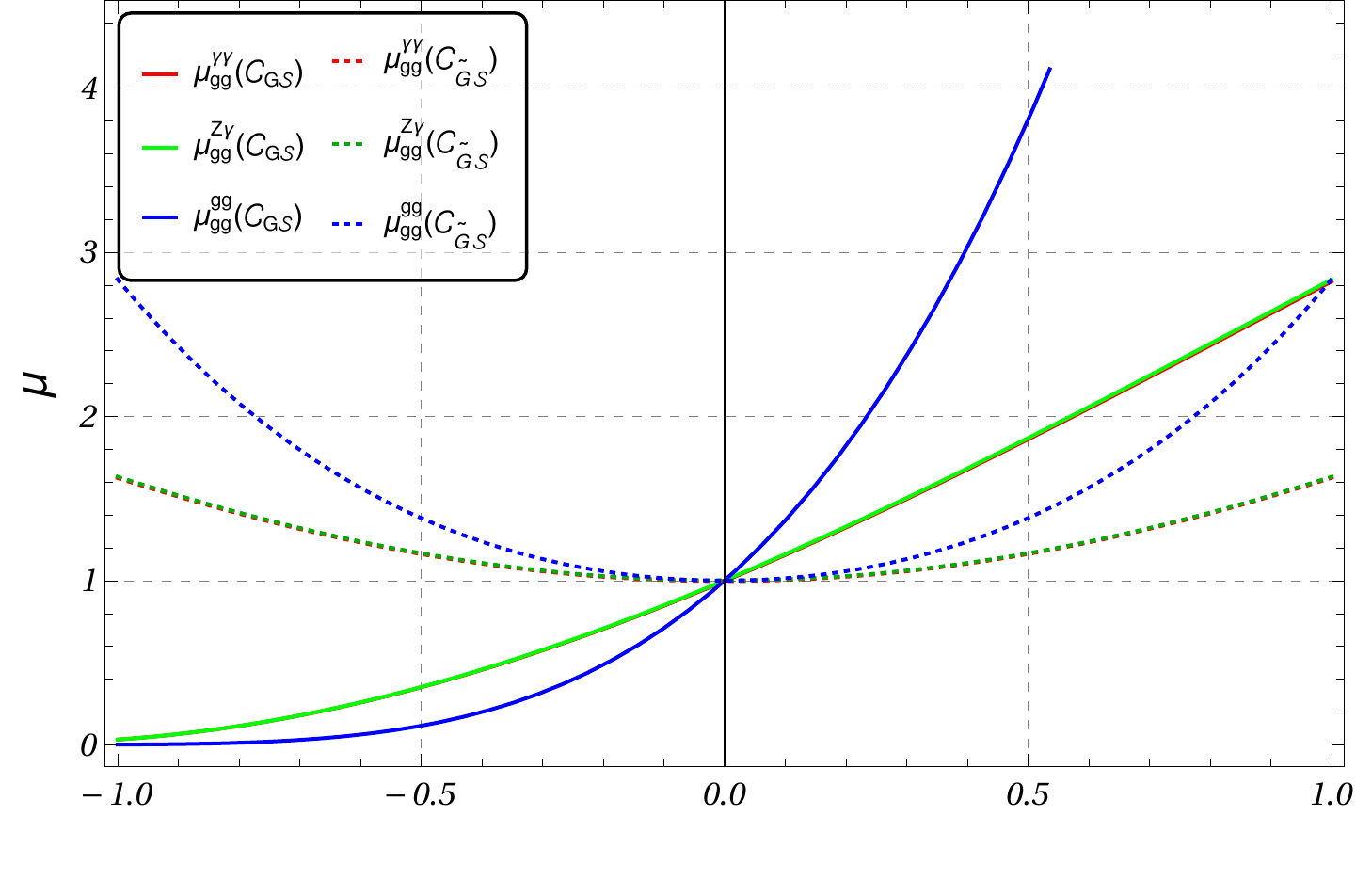}} %$C_{GS}, \widetilde C_{\widetilde{G}S}$
	\caption{Higgs boson signal strength modifications due to the different $h^\pm$-related effective operators as a function of their Wilson coefficients for $M_h=700~\text{GeV}$ and $\Lambda=1~\text{TeV}$. The CP-even operators are well-approximated by a linearised calculation for perturbative choices while the CP-odd operators impact the inclusive Higgs properties at squared dimension 6 level by construction. The gauge-$h^\pm$ operators (a,c,d) have a more significant impact on the Higgs phenomenology than the Higgs-$h^\pm$ operators (b). The effective interactions related to the gluon are given in (d), but we will mainly focus on electroweak couplings in this work (see text).
	\label{fig:signalstrength}}
\end{figure*}
%%%%%%%%%%%%%%%%%%%%%%%%%%%%%%%%%%

%%%%%%%%%%%%%%%%%%%%%%%%%%%%%%%%%%
\begin{figure}[t]
	\includegraphics[width=0.45\textwidth]{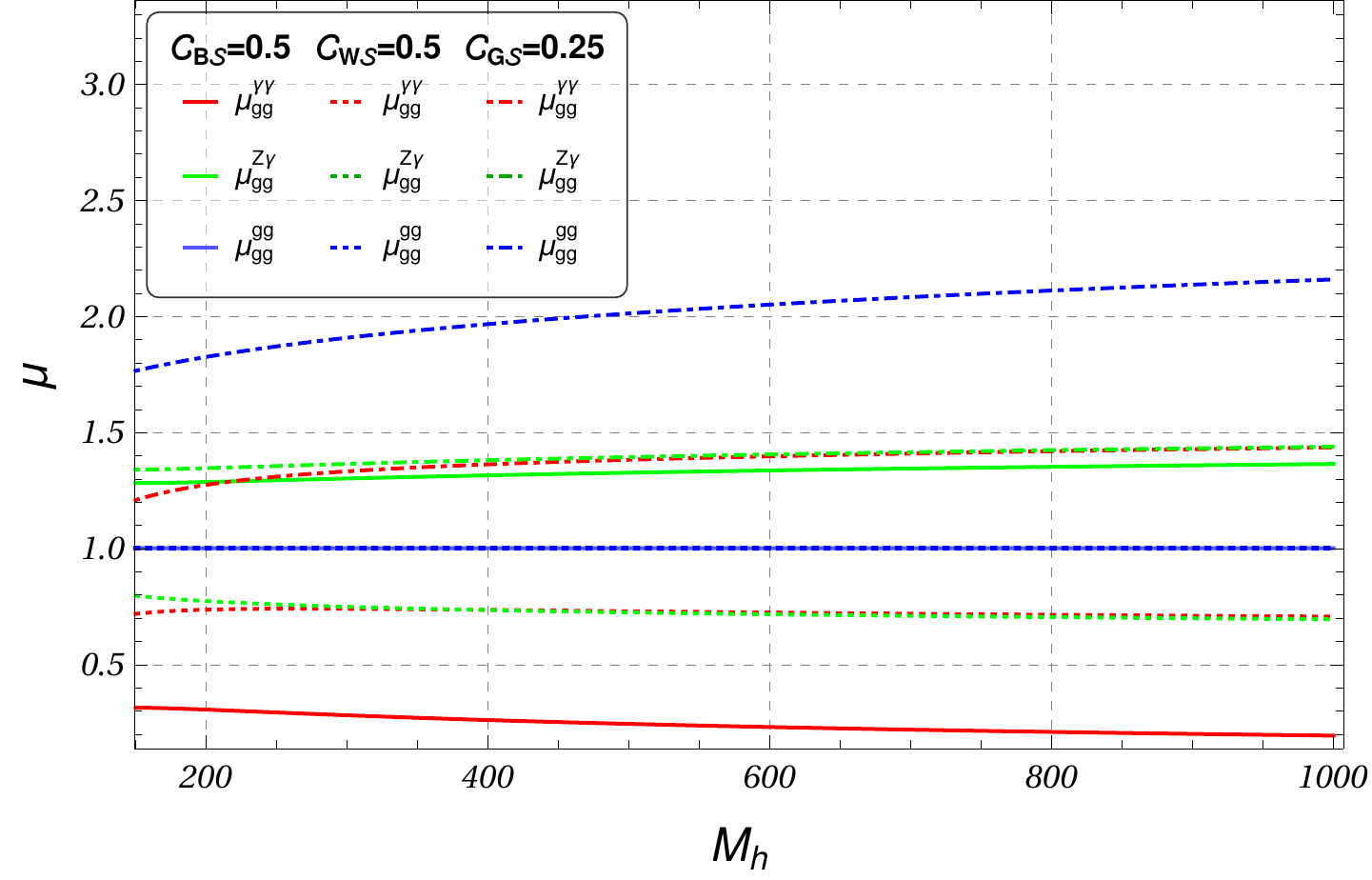}
	\caption{Impact of the $\mathcal{C}_{W\mathcal{S}},\mathcal{C}_{B\mathcal{S}},\mathcal{C}_{G\mathcal{S}}$ operators on the Higgs signal strength in the different loop-induced decay modes as a function of the new scalar mass $M_h$. \label{fig:signalstrength2}}
\end{figure}
%%%%%%%%%%%%%%%%%%%%%%%%%%%%%%%%%%

\subsubsection*{Constraints}

Before we will discuss the impact of the considered scenario on the SM Higgs boson's phenomenology as outlined above, a few remarks regarding constraints on the model are due. 

We have already commented on the potential lepton flavour implications in Sec.~\ref{sec:model}, in the limit where interactions of $h^\pm$ with fermions is weak, flavour constraints can be avoided. Yet, resonantly-produced $h^\pm$ can still be observed in leptonic final states ($\text{BR}\simeq 1$) and searches for charged Higgs bosons at the LHC~(e.g. \cite{Aaboud:2018cwk,Sirunyan:2019arl,ATLAS:2020jqj,Sirunyan:2020aln,Sirunyan:2020hwv}) typically focus on their quark or lepton decay final states. However, searches for gauge-philic charged scalars with suppressed the Yukawa interactions have limited sensitivity, see e.g.~\cite{Liu:2013oen,Adhikary:2020cli}. The recent Ref.~\cite{Sirunyan:2017sbn} (see also~\cite{Aad:2015nfa}) sets 95\% confidence level constraint on a charged Higgs in decays $h^\pm\to W^\pm Z$ between 90~fb and 1~pb for masses in the range of $0.4~\text{TeV} \lesssim M_h \lesssim 2~\text{TeV}$. However, these final states exploit a non-trivial role of $h^\pm$ in electroweak symmetry breaking (as part of e.g. a $SU(2)_L$ triplet~\cite{Georgi:1985nv}), and hence rest on the assumption of a significant departure of the alignment of $H$ from fluctuations around $v$, and a significant non-doublet character of SSB. The charged Higgs bosons introduced above are produced at the LHC via Drell-Yan-like pair production (e.g., as also present in the two-Higgs-doublet model). While even parametrically small Yukawa couplings can lead to discoverable clean leptonic final states as mentioned above, the electroweak pair production cross-section is suppressed such that the LHC will be statistically limited in a mass range $M_h\simeq 500~\text{GeV}$ (see also~\cite{ Englert:2016ktc}). It is worth noting that the gauge-$h^\pm$ effective field theory insertions do not lead to an enhancement of Drell-Yan production at large energies.

Next, we comment on constraints on the new couplings from unitarity and perturbativity. The scattering angle of $2\to 2$ scattering can be removed by projecting the amplitude on to partial waves
\begin{multline}
\label{eq:unit}
a^J_{fi} ={\beta^{1/4}(s,m_{f1}^2,m_{f2}^2) \beta^{1/4}(s,m_{i1}^2,m_{i2}^2)\over 32\pi s} \\ \int_{-1}^{1} \hbox{d} \cos\theta \,D^J_{\mu_i,\mu_f} \, i{\cal{M}}_{fi} \left(\cos\theta, \sqrt{s}\right),
\end{multline} 
where $s$ is squared the centre-mass-energy and the $m_i$ are the masses of the states in the initial state $i$ and final state $f$. $i{\cal{M}}$ is the $2\to 2$ scattering amplitude (identical particles in initial and final states require an additional factor $1/\sqrt{2}$), $D^J_{\mu_i,\mu_j}$ are the Wigner functions of \cite{Jacob:1959at}, $\mu_{i,j}$ are defined from the differences of the initial and final states helicities (see also \cite{DiLuzio:2016sur}), and $\beta(x,y,z)=x^2+y^2+z^2-2xy-2xz-2yz$. Unitarity and perturbativity can then be parametrised as $|a^J_{fi}|<1$~\cite{Lee:1977yc,Lee:1977eg,Chanowitz:1978uj,Chanowitz:1978mv}. 

%%%%%%%%%%%%%%%%%%%%%%%%
\begin{table}[!t]
	\centering
	\renewcommand{\arraystretch}{1.9}
	{\scriptsize\begin{tabular}{||c|c||c|c||}
			\hline
			\hline
			\multicolumn{2}{||c||}{$\Phi^4\mathcal{D}^2$}&
			\multicolumn{2}{c||}{$\Phi^6$}\\
			\hline
			
			$\mathcal{O}_{ \mathcal{S} \phi\mathcal{D}} $&
			$(\mathcal{S}^{\dagger}\,\mathcal{S})\,\left[(\mathcal{D}^{\mu}\,\phi)^{\dagger}(\mathcal{D}_{\mu}\,\phi)\right]$&
			$\mathcal{O}_{\mathcal{S}\phi} $&
			$(\phi^{\dagger} \,\phi) \,(\mathcal{S}^{\dagger} \,\mathcal{S})^2$\\
			
			$\mathcal{O}_{\phi \mathcal{S} \mathcal{D}}$&
			$(\phi^{\dagger}\,\phi)\,\left[(\mathcal{D}^{\mu}\,\mathcal{S})^{\dagger}(\mathcal{D}_{\mu}\,\mathcal{S})\right]$&
			&
			\\
			
			$\mathcal{O}_{ \mathcal{S} \Box}$&
			$(\mathcal{S}^{\dagger}\,\mathcal{S})\,\Box\,(\mathcal{S}^{\dagger}\,\mathcal{S})$&
			&
			\\
			\hline
			\hline
			\multicolumn{4}{||c||}{$\Phi^2X^2$}\\
			\hline
			
			$\mathcal{O}_{ B \mathcal{S}} $&
			$B_{\mu\nu} \,B^{\mu\nu} \,(\mathcal{S}^{\dagger} \,\mathcal{S})$&$\mathcal{O}_{\widetilde{B} \mathcal{S}} $
			&$\widetilde{B}_{\mu\nu} \,B^{\mu\nu} \,(\mathcal{S}^{\dagger} \,\mathcal{S})$
			\\
			
			$\mathcal{O}_{ W \mathcal{S}} $&
			$W^{I}_{\mu\nu} \,W^{I\mu\nu} \,(\mathcal{S}^{\dagger} \,\mathcal{S})$&
			$\mathcal{O}_{ \widetilde{W} \mathcal{S} } $&
			$\widetilde{W}^{I}_{\mu\nu} \,W^{I\mu\nu} \,(\mathcal{S}^{\dagger} \,\mathcal{S})$
			\\
			\hline
	\end{tabular}}
	\caption{Effective operators contributing to $2 \to 2$ scattering amplitudes. The operators from $\Phi^4\mathcal{D}^2$ and $\Phi^6$ classes contribute to $h^\pm h^\mp$ scattering while $\Phi^2X^2$ operators affect $h^\pm W^\mp$, $h^\pm \gamma$ and $h^\pm Z$ scattering. }
	\label{table:unitarity}
\end{table}
%%%%%%%%%%%%%%%%%%%%%%%%

$h^\pm h^\mp$ scattering receives non-negligible corrections from the effective field theory operators $\mathcal{O}_{\mathcal{S}\Box}, \mathcal{O}_{\phi \mathcal{S} \mathcal{D}}, \mathcal{O}_{\mathcal{S}\phi}$ mentioned in Table~\ref{table:unitarity}, in the high energy regime $\sqrt{s}\gg M_h^2$ (and other contributing mass scales). Therefore, perturbativity of the $J=0$ partial wave can be used to restrict the Wilson coefficient range at dimension 6 level
\begin{equation}
\label{eq:pert1}
\begin{split}
{|{\cal{C}}_{\phi \mathcal{S} D} |\over \Lambda^2}  & \lesssim {32\pi \over |2\lambda_2- \lambda_3| v^2} \,,\\
{|{\cal{C}}_{\mathcal{S}\phi } |\over \Lambda^2}  & \lesssim {16\pi \over v^2} \,,\\
{|{\cal{C}}_{\mathcal{S}\Box} |\over \Lambda^2}  & \lesssim {32\pi \over s}\,, \\
{|{\cal{C}}_{\mathcal{S}\phi D} |\over \Lambda^2}  & \lesssim {64\pi \over s} \,.\\
\end{split}
\end{equation}
while ${\cal{O}}_{\phi \mathcal{S}}$ has a vanishing contribution in this kinematic regime. $\lambda_{2,3}$ as renormalisable interactions are subject to the usual $\sim 4\pi$ bound.
We will see that the operators of Eq.~\eqref{eq:pert1} only have a mild impact on the Higgs phenomenology below. Even non-perturbative coupling choices $\sim 4\pi/\text{TeV}^2$ do not lead to phenomenologically relevant deviations. The gauge scalar-operators are more relevant for driving the BSM Higgs physics modifications (see below) and can be analysed by considering $h^\pm W^\mp$, $h^\pm \gamma$ and $h^\pm Z$ scattering. Using the strategy of \cite{Chanowitz:1978uj,Chanowitz:1978mv} we compute coupled $J=1$ bounds (note that in for effective interactions transverse $V$ polarisations provide constraints which is qualitatively different from the SM~\cite{Englert:2016ljt})
\begin{equation}
\begin{split}
{|{\mathcal{C}}_{W\mathcal{S}} |\over \Lambda^2}, {|{\mathcal{C}}_{\widetilde{W} \mathcal{S}} |\over \Lambda^2} &\lesssim {24\pi\over M_W \sqrt{s}}  \,,\\
{|{\cal{C}}_{B\mathcal{S}} |\over \Lambda^2}, {|{\cal{C}}_{\widetilde{B} \mathcal{S}} |\over \Lambda^2} &\lesssim  {48\pi\over M_W c_\theta^2 \sqrt{s} },
\end{split}
\end{equation}
again in the limit where participating masses are negligible compared to $\sqrt{s}$. These limits are rather weak, for example unitarity violation at $\sqrt{s}\simeq 10~\text{TeV}$ translates into rather loose bounds of $|{\cal{C}}_i | \lesssim 100~{\Lambda^2/\text{TeV}^2}$.

Thirdly, one might object at this point that electroweak precision measurements such as the oblique corrections already constrain this scenario. To clarify this point, we have investigated the Peskin-Takeuchi $S,T,U$ parameters~\cite{Peskin:1990zt,Peskin:1991sw} in the scenario of Sec.~\ref{sec:model} where the gauge boson polarisations receive ${\cal{C}}_{W\mathcal{S}}$ and ${\cal{C}}_{B\mathcal{S}}$-related corrections. We find that these Wilson coefficients identically vanish from the on-shell renormalised $S,T,U$ parameters leaving a residual dependence on the mass scale $M_h$. However, given that these states are weakly coupled, their contribution is small to the extent that this scenario is not constrained by electroweak precision data. 

%%%%%%%%%%%%%%%%%%%%%%%%%%
\begin{figure}[t!]
	\includegraphics[width=0.49\textwidth]{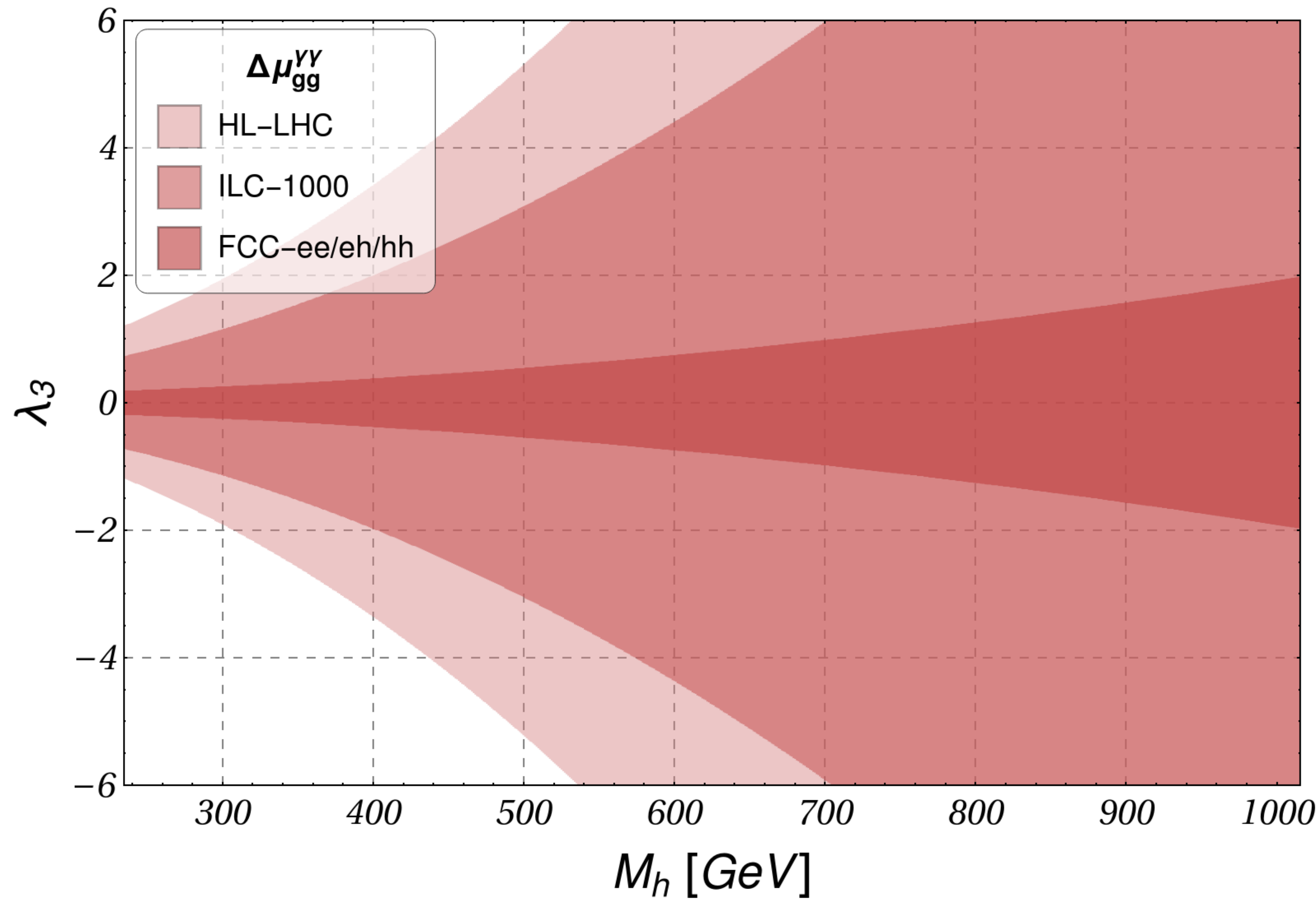}
	\caption{Limits of the projected $H\to \gamma\gamma$ signal strength at 3/ab HL-LHC~($\Delta\mu=3.3\%$)~\cite{CMS:2017cwx}, the ILC-1000 ($\Delta\mu=1.9\%$) and the FCC-ee/eh/hh~($\Delta\mu=0.29\%$)~\cite{deBlas:2019rxi}. No dimension 6 effects are included. \label{fig:hgamgam}}
\end{figure}
%%%%%%%%%%%%%%%%%%%%%%%%%%
%%%%%%%%%%%%%%%%%%%%%%%%%%
\begin{figure*}[!t]
	\subfigure[\label{fig:hcanc1}]{\includegraphics[width=0.49\textwidth]{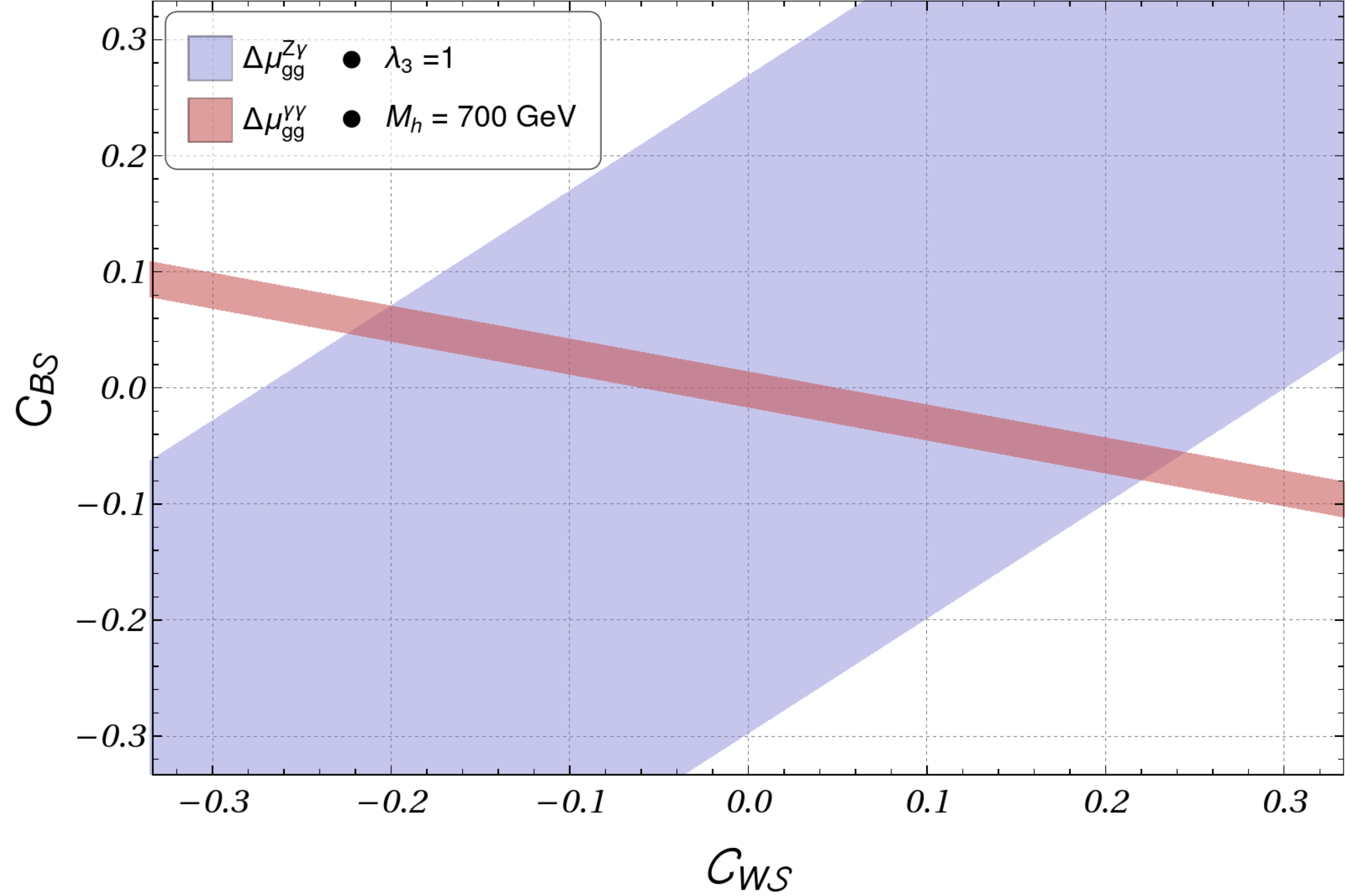}}
	\hfill
	\subfigure[\label{fig:hcanc2}]{\includegraphics[width=0.49\textwidth]{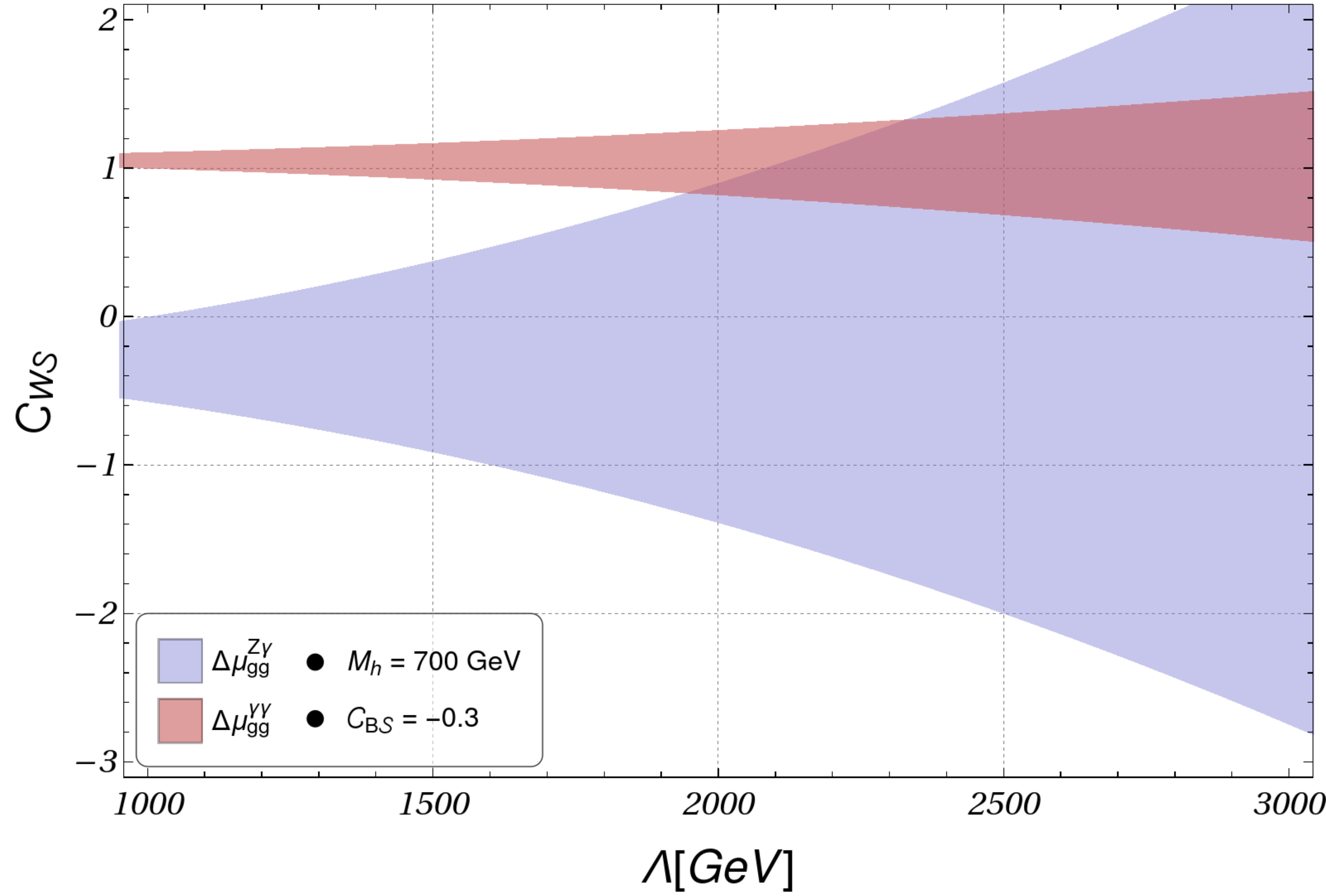}}
	\caption{Complementarity of the ${\cal{C}}_{B\mathcal{S}},{\cal{C}}_{W\mathcal{S}}$ directions in the $H\to \gamma \gamma,Z\gamma$ comparison. If one of the operators is present but due to a cancellation, the $H\to \gamma\gamma$ rate looks SM-compatible, the $Z\gamma$ final state can resolve this blind direction.\label{fig:hcanc}}
\end{figure*}
%%%%%%%%%%%%%%%%%%%%%%%%%%

\medskip
Under the assumptions of this work, namely that new physics contributions are predominantly mediated through the charged Higgs sector, and its non-negligible interactions with the SM Higgs sector, the precision investigation of the $H\to \gamma\gamma, Z\gamma$ decay as outlined above can act as an indirect and phenomenologically important probe of such extensions. 

\subsubsection*{Loop-induced Higgs phenomenology}

Returning to loop-induced Higgs boson decays, we present the signal strength deviations from the SM as a function of a range of Wilson coefficients of Sec.~\ref{sec:model} in Fig.~\ref{fig:signalstrength}, for $M_h=700~\text{GeV}$, $\Lambda=1~\text{TeV}$, and $\lambda_3=1$, as well as vanishing $\overline{\text{MS}}$ values for the couplings of Eq.~\eqref{eq:bare}. As indicated earlier, the pseudo-observables are well-described by the linearised approximation for perturbative choices of the Wilson coefficients.  The obvious exceptions are the CP-odd interactions where the interference of CP-even SM amplitude and dimension 6 CP-odd contributions cancels identically in CP-even observables like the partial decay widths. CP-even effects then arise as squared CP-odd dimension 6 contributions, giving rise to a non-linear Wilson coefficient dependence. 

Furthermore, we note that the effect of electroweak corrections to $\mu^{gg}_{gg}$ is negligible for $\mathcal{C}_{G\phi,\mathcal{S}},\mathcal{C}_{\tilde{G}\phi,\mathcal{S}}=0$ and results from the small overall modification of the Higgs total decay width, Fig.~\ref{fig:signalstrength2}. This is the limit where our results are most relevant: BSM degrees of freedom with non-trivial QCD interactions that are integrated out to arrive at $\mathcal{C}_{G\phi,\mathcal{S}},\mathcal{C}_{\tilde{G}\phi,\mathcal{S}}\neq 0$ can typically can be more efficiently constrained by direct searches at hadron colliders, see e.g. the recent discussion of~\cite{Englert:2019rga,Brown:2020uwk}.

Turning to the effective electroweak interactions, the $Z\gamma$ and $\gamma\gamma$ decay widths are particularly sensitive to modifications of the gauge-$h^\pm$ interaction for the chosen Wilson coefficient normalisations, while the ${\cal{C}}_{\phi \mathcal{S}},{\cal{C}}_{\phi \mathcal{S} \mathcal{D}}$ interactions are suppressed. Phenomenologically relevant deviations from the SM expectations related to ${\cal{C}}_{\phi \mathcal{S}},{\cal{C}}_{\phi \mathcal{S} \mathcal{D}}$ are quickly pushed to the non-perturbative coupling regime where a meaningful perturbative matching is not possible. This indicates a phenomenological blindness of Higgs signal strength data to the interactions parametrised by these coefficients, also because of gauge cancellations between the diagrams of Fig.~\ref{fig:feyn}.

In Fig.~\ref{fig:hgamgam}, we show the expected constraints charged Higgs masses as a function of the coupling $\lambda_3$, for $\Lambda =\infty$. The LHC will be able to indirectly probe the gauge-philic scenario up mass scales of $\sim 500~\text{GeV}$ for perturbative scenarios, while sensitivity extrapolations at the FCC-hh~\cite{Benedikt:2018csr} and the highly constraining FCC-ee can explore a broader range of charged Higgs bosons. The inclusion of higher-dimensional interactions related to the new charged scalar changes this picture. 

We will first focus on the expected outcome of the HL-LHC. Extrapolations by the CMS experiment~\cite{CMS:2017cwx} suggest that
\begin{equation}
{\Delta\mu^{\gamma\gamma}_{gg} \over \mu^{\gamma\gamma}_{gg}} = 3.3\%,
\end{equation}
can be obtained at a luminosity of 3/ab. The $H\to Z\gamma$ is considerably more challenging and statistically limited in the recent 139/fb ATLAS analysis of \cite{Aad:2020plj} which gives an expected $\mu^{Z\gamma} = 1.0 \pm  0.8~\text{(stat.)} \pm 0.3~\text{(syst.)}$. Rescaling uncertainties with the root of the luminosity, we can estimate the sensitivity at 3/ab to be
\begin{equation}
{\Delta\mu^{Z\gamma}_{gg}\over\mu^{Z\gamma}_{gg}} = 18\%\,,
\end{equation} 
which is comparable with the extrapolation of \cite{deBlas:2019rxi} in the context of the $\kappa$ framework~\cite{Dittmaier:2011ti}.
Furthermore, extrapolating to a future circular collider, Ref.~\cite{deBlas:2019rxi} quotes improvements of
\begin{equation}
{\Delta\mu^{\gamma \gamma}_{gg}\over \mu^{\gamma\gamma}_{gg}} = 0.6\%\,,~
{\Delta\mu^{Z\gamma}_{gg}\over \mu^{Z\gamma}_{gg}} = 1.4\%,
\end{equation}
from combinations of the ee, eh, and hh options~\cite{,Benedikt:2018csr,Abada:2019lih,Abada:2019zxq}.

The $Z\gamma$ and $\gamma\gamma$ channels access orthogonal information of the dimension six interactions, Fig.~\ref{fig:hcanc1}. For an SM-like outcome of both $H$ decay channel measurements within the uncertainty quoted above, the ${\cal{C}}_{W\mathcal{S}}$ and ${\cal{C}}_{B\mathcal{S}}$ operators yield complementary constraints as a result of different overlaps of $Z,\gamma$ with the gauge eigenstates. Concretely this means that if one of the operators is expected to be non-zero, the combination of both channels can be used as a measurement or constraint on other contributing effective couplings as demonstrated in Fig.~\ref{fig:hcanc2} for the case of ${\cal{C}}_{W\mathcal{S}}/\Lambda^2$.

%%%%%%%%%%%%%%%%%%%%%%%%%%%%%%%%%%
\section{Conclusions}\label{sec:conc}
%%%%%%%%%%%%%%%%%%%%%%%%%%%%%%%%%%
The presence of additional charged scalar degrees of freedom is predicted in many BSM scenarios. When these states couple predominantly to the electroweak sector, they are difficult to observe experimentally, in particular when they do not play a role in electroweak symmetry breaking. This highlights the question of whether additional new physics that arises as a non-trivial extension of the extra scalar's interactions can have phenomenologically relevant implications. 

We approach this question by means of effective field theory, i.e. we assume a mass gap between the charged BSM scalar and other states that lead to generic effective operators involving the scalar and the Standard Model fields. While in the most generic approach, all SMEFT operators would be sourced as well, these can be radiative effects when the new states predominantly interact with the SM fields via the propagating scalar (as also motivated, e.g. from Higgs portals). 

While precision electroweak observables are largely unaffected by the presence of this state, loop-induced Higgs decays become sensitive tools to set constraints for these (strong) new physics contributions associated with the charged scalar. In particular, operator combinations that are not constrained by generic gauge boson phenomenology can be accessed in a precision analysis of Higgs decays into rare yet clean $\gamma\gamma$ and $\gamma Z$ final states. As we have demonstrated, the complementarity of these decay modes could allow us, at least to some extent, to disentangle new physics contributions in case this scenario is broadly realised.

\acknowledgements
The work of A, U.B., and J.C. is supported by the Science and Engineering Research Board, Government of
India, under the agreements SERB/PHY/2016348 and SERB/PHY/2019501 and Initiation Research
Grant, agreement number IITK/PHY/2015077, by IIT Kanpur. C.E. is supported by the UK Science and Technology Facilities Council (STFC) under grant ST/T000945/1 and by the IPPP Associateship Scheme. M.S. is supported by the STFC under grant ST/P001246/1.

\appendix

\appendix
\allowdisplaybreaks
\section{Renormalisation}\label{sec:appendix}

As discussed in Sec.~\ref{sec:calc}, we have considered on-shell renormalisation for the SM and additional fields and parameters, and $\overline{\text{MS}}$ renormalisation for Wilson coefficients. Here we have given the explicit expressions for the renormalisation constants used in the counter term given in Eq.~\eqref{eq:cts}. The terms $A_{0},\,B_{0},\,B_{00},\,B_{1}$ used in the following equations are the short-handed notations to express the one-point and two-point integrals (see e.g.~\cite{Denner:1991kt})
\begin{equation}
\begin{split}
A_{0}(m^2)&=m^2 \,\Delta+\mathcal{O}(1),\nonumber\\
B_{0}&=\Delta+\mathcal{O}(1),\nonumber\\
B_{1}&=-\frac{\Delta}{2}+\mathcal{O}(1),\nonumber\\
B_{00}(p^2,m_{1}^2,m_{2}^2)&=\left(\frac{m_1^2+m_2^2}{4}-\frac{p^2}{12}\right)\Delta+\mathcal{O}(1).
\end{split}
\end{equation}
$\Delta\sim \epsilon^{-1}$ denotes the UV-divergent $\overline{\text{MS}}$ parts of the one-loop integrals in dimensional regularisation $d=4-2\epsilon$.

The wave function renormalisation are computed from the on-shell conditions of the two-point functions at $p^2=0$,
\begin{alignat}{4}
\delta Z_{AA}&=\frac{1}{96 \pi ^2 \left(g_{_Y}^2+g_{_W}^2\right)}\Big[g_{_Y}^2 g_{_W}^2 \Big(4+30 B_{0}(M_{_W}^2)\nonumber\\
& +24
\sum_{l=e,\mu,\tau} B_{1}(M_{l}^2)+8 \sum_{q_d=d,s,b}B_{1}(M_{q_d}^2)\nonumber\\
&+32
\sum_{q_u=u,c,t}B_{1}(M_{q_u}^2)-48\sum_{l=e,\mu,\tau}
dB_{00}(M_{l}^2)\nonumber\\
&-16\sum_{q_d=d,s,b}
dB_{00}(M_{q_d}^2)-64\sum_{q_u=u,c,t}
dB_{00}(M_{q_u}^2)\nonumber\\
&+12 B_{1}(M_{_W}^2)-3 \,g_{_W}^2 v^2
dB_{0}(M_{_W}^2)+12 M_{_W}^2 dB_{0}(M_{_W}^2)\nonumber\\
&+12
dB_{00}(M_{_W}^2)+24 dB_{00}(M_{h}^2)+60 dB_{00}(M_{_W}^2)\Big)\nonumber\\
&-24
A_{0}(M_{h}^{2}) \left(\mathcal{C}_{B\mathcal{S}} \,g_{_W}^2+\mathcal{C}_{W\mathcal{S}} \,g_{_Y}^2\right)\Big],	
\end{alignat}
and
\begin{alignat}{5}
\delta Z_{ZA}&=-\frac{1}{48 \pi ^2 M_{_Z}^2 (g_{_Y}^2+g_{_W}^2)}\Big[g_{_Y} g_{_W} \Big(3 g_{_Y}^2 g_{_W}^2 v^2 B_{0}(M_{_W}^2)\nonumber\\
&+12 g_{_W}^2
M_{_W}^2 B_{0}(M_{_W}^2)-12 g_{_Y}^2 B_{00}(M_{_W}^2)\nonumber\\
&-24 g_{_W}^2 A_{0}(M_{_W}^2)-24
g_{_Y}^2 B_{00}(M_{h}^2)\nonumber\\
&+(36 g_{_Y}^2 - 12 g_{_W}^2)\sum_{l=e,\mu,\tau}
B_{00}(M_{l}^2)+(4 g_{_Y}^2-12 g_{_W}^2)\nonumber\\
& \sum_{q_d=d,s,b}
B_{00}(M_{q_d}^2)+(40 g_{_Y}^2-24g_{_W}^2)\sum_{q_u=u,c,t}
B_{00}(M_{q_u}^2)\nonumber\\
&+60 g_{_W}^2
B_{00}(M_{_W}^2)+6
\left(g_{_Y}^2-g_{_W}^2\right)A_{0}(M_{_W}^2)\nonumber\\
&+12 g_{_Y}^2 A_{0}(M_{h}^2)+(6 g_{_W}^2-18 g_{_Y}^2) \sum_{l=e,\mu,\tau}A_{0}(M_{l}^2)\nonumber\\
&+(6 g_{W}^2-2 g_{_Y}^2)\sum_{q_d=d,s,b} A_{0}(M_{q_d}^2)+(12 g_{_W}^2-20 g_{_Y}^2)\nonumber\\
& \sum_{q_u=u,c,t}A_{0}(M_{q_u}^2)\Big)\Big]\,.\nonumber\\
\end{alignat}
Note that the dimension 6 parts of these renormalisation constants would introduce dimension eight contributions which we neglect consistently in the computation of the next-to-leading order dimension six amplitude (see also~\cite{Grojean:2013kd,Englert:2014cva,Englert:2019rga}.

Similarly, for the Higgs boson, the wave function renormalisation is computed from an on-shell residue at $p^2=m_H^2$ (thus eliminating LSZ factors from the $S$-matrix element) 
\begin{alignat}{5}
\delta Z_{H}&=	\frac{1}{64 \pi ^2}\Big(\mathcal{C}_{\phi\mathcal{S}\mathcal{D}} \,\lambda_{3}^2 \,v^4 \,dB_{0}(M_{H}^2,M_{h}^2) \nonumber\\
&+4 \,\mathcal{C}_{\phi\mathcal{S}}
\,\lambda_{3}\,v^4 \,dB_{0}(M_{H}^2,M_{h}^2)+4 \,v^2 \,\mathcal{C}_{\phi\mathcal{S}\mathcal{D}} \,\lambda_{3}
B_{1}(M_{H}^2,M_{h}^2)\nonumber\\
&+4 \,\mathcal{C}_{\phi\mathcal{S}\mathcal{D}} \,\lambda_{3} \,M_{h}^2 \,v^2
\,dB_{0}(M_{H}^2,M_{h}^2)\nonumber\\
&+4 \,\mathcal{C}_{\phi\mathcal{S}\mathcal{D}} \,\lambda_{3} M_{H}^2 \,v^2
\,dB_{1}(M_{H}^2,M_{h}^2)+4 \,\mathcal{C}_{\mathcal{S}\phi\mathcal{D}}
A_{0}(M_{h}^2)\Big)\nonumber\\
&+\frac{1}{256 \,\pi^2 \,v^2}\Big(128\sum_{l=e,\mu,\tau}\,M_{l}^4
B_{0}(M_{H}^2,M_{l}^2)  \nonumber\\
&+64\sum_{l=e,\mu,\tau}\,M_{l}^2
B_{1}(M_{H}^2,M_{l}^2) +64 M_{H}^2\nonumber\\
&
\sum_{l=e,\mu,\tau}dB_{1}(M_{H}^2,M_{l}^2) \,M_{l}^2+12 (g_{_Y}^2+g_{_W}^2)v^2 B_{0}(M_{H}^2,M_{Z}^2)
\nonumber\\
&+24 g_{_W}^2 \,v^2
B_{0}(M_{H}^2,M_{W}^2)+192 \sum_{q_d=d,s,b} \,m_{q_d}^2
B_{1}(M_{H}^2,M_{q_d}^2)\nonumber\\
&+192 \sum_{q_u=u,c,t} \,M_{q_u}^2
B_{1}(M_{H}^2,M_{q_u}^2)\nonumber\\
&+384\sum_{q_d=d,s,b}\,M_{q_d}^4
dB_{0}(M_{H}^2,M_{q_d}^2)\nonumber\\
&+384 \sum_{q_u=u,c,t}M_{q_u}^4 dB_{0}(M_{H}^2,M_{q_u}^2)\nonumber\\
&+384 \sum_{q_d=d,s,b}\,M_{q_d}^4dB_{0}(M_{H}^2,M_{q_d}^2)\nonumber\\
&+192 \,M_{H}^2 \sum_{q_d=d,s,b}M_{q_d}^2
\,dB_{1}(M_{H}^2,M_{q_d}^2)\nonumber\\
&+192 \,M_{H}^2 \sum_{q_u=u,c,t}M_{q_u}^2
\,dB_{1}(M_{H}^2,M_{q_u}^2)\nonumber\\
&+16 g_{_W}^2 v^2
B_{1}(M_{H}^2,M_{W}^2)+8 (g_{_Y}^2+g_{_W}^2) v^2
B_{1}(M_{H}^2,M_{Z}^2)\nonumber\\
&-8 \lambda^2 v^4
dB_{0}(M_{H}^2,M_{Z}^2)+4 (g_{_Y}^2+g_{_W}^2) M_{Z}^2 v^2
dB_{0}(M_{H}^2,M_{Z}^2)
\nonumber\\
&+12 (g_{_Y}^2+g_{_W}^2)\,M_{H}^2 v^2
dB_{0}(M_{H}^2,M_{Z}^2)\nonumber\\
&+12 g_{_W}^2 M_{H}^2 v^2
dB_{0}(M_{H}^2,M_{Z}^2)-16 \lambda_{1}^2 v^4
dB_{0}(M_{H}^2,M_{W}^2)\nonumber\\
&+8 g_{_W}^2 M_{W}^2 v^2
dB_{0}(M_{H}^2,M_{W}^2)+24 g_{_W}^2 M_{H}^2 v^2\nonumber\\
&
dB_{0}(M_{H}^2,M_{W}^2)-14 g_{_W}^4 v^4
dB_{0}(M_{H}^2,M_{W}^2)\nonumber\\
&-7(g_{_Y}^4+g_{_W}^4) v^4
dB_{0}(M_{H}^2,M_{Z}^2)-14 g_{_W}^2 g_{_Y}^2 v^4
\nonumber\\ 
&dB_{0}(M_{H}^2,M_{Z}^2)-4 \lambda_{3} v^4
dB_{0}(M_{H}^2,M_{h}^2)-72 \,\lambda_{1}^2 \,v^4
\nonumber\\
&dB_{0}(M_{H}^2)
+16 g_{_W}^2 M_{H}^2 v^2
\,dB_{1}(M_{H}^2,M_{_W}^2)+8 (g_{_Y}^2+g_{_W}^2)\nonumber\\
& M_{H}^2 v^2
\,dB_{1}(M_{H}^2,M_{Z}^2)\Big)\,,
\end{alignat}
where $dB_i$ represents the derivative of the scalar function with respect to the external momentum. The tadpole counter term $\delta v$ in Eq.~\eqref{eq:cts} reads,

\begin{alignat}{6}
\delta v &=-\frac{1}{64 \pi ^2 M_{H}^2 v}\Big[v^2 A_{0}(M_{h}^2) \Big(4 \,v^2 \,\mathcal{C}_{\phi\mathcal{S}} +\lambda_{3} \,v^2 \,\mathcal{C}_{\phi\mathcal{S}} \nonumber\\
&+4 \,M_{h}^2 \,\mathcal{C}_{\phi\mathcal{S}\mathcal{D}}
-2 \lambda_{3}\Big)-3 \,g_{_Y}^2 v^2 A_{0}(M_{Z}^2)-6 g_{_W}^2 v^2 A_{0}(M_{W}^2)\nonumber\\
&-3 g_{_W}^2 \,v^2 A_{0}(M_{Z}^2)-2 \lambda_{1} v^2 A_{0}(M_{Z}^2)-6 \lambda_{1} v^2
A_{0}(M_{H}^2)\nonumber\\
&-4 \lambda_{1} v^2 A_{0}(M_{W}^2)+16 \sum_{l=e,\mu,\tau}M_{l}^2 A_{0}(M_{l}^2)\nonumber\\
&+48 \sum_{q_d=d,s,b}M_{q_d}^2
A_{0}(M_{q_d}^2)+48 \sum_{q_u=u,c,t}M_{q_u}
A_{0}(M_{q_u}^2)\nonumber\\
&+2 (g_{_Y}^2+g_{_W}^2) v^2 M_{Z}^2 +4
g_{_W}^2 v^2 M_{W}^2\Big].
\end{alignat}

These terms need to be included in the renormalisation of the three-point function $H\to \gamma\gamma$ of Fig.~\ref{fig:feyn}. The divergences related to the renormalisation of the Wilson coefficients are then given by
\begin{alignat}{4}
\delta\mathcal{C}_{A\phi} & =  \frac{4 g_{_Y}g_{_W}^3M_{H}^2M_{_W}^2+g_{_Y}^3g_{_W}^3M_{H}^2v^2}{64(g_{_Y}^2+g_{_W}^2)M_{H}^2M_{Z}^{2}\pi^2}\,\mathcal{C}_{AZ\phi}\nonumber \\ & +\frac{\lambda_{3}\,(g_{_Y}^2\,\mathcal{C}_{W\mathcal{S}}+g_{_W}^2\,\mathcal{C}_{B\mathcal{S}})}{32(g_{_Y}^2+g_{_W}^2)\pi^2}\nonumber \\ &  
+\frac{1}{64 \pi ^2 M_{H}^2 v^2
	(g_{_Y}^2+g_{_W}^2)}\Big[2 g_{_Y}^2 \Big(v^2 (-(g_{_W}^2 \nonumber\\
&  (3(M_{W}^2+M_{Z}^2)-14 M_{H}^2)+\lambda_{1} (3 M_{H}^2+2
M_{W}^2\nonumber\\
& +M_{Z}^2)+\lambda_{3} M_{h}^2))+2 M_{H}^2 (\sum_{l=e,\mu,\tau}M_{l}^2 +3
\sum_{q=u,d}M_{q}^2)\nonumber\\
& +8 (\sum_{l=e,\mu,\tau}M_{l}^4+3
\sum_{q=u,d}M_{q}^4)\Big)-g_{_Y}^4 v^2 (M_{H}^2+3
M_{Z}^2)\nonumber\\
& +g_{_W}^2 (4 (M_{H}^2 (\sum_{l=e,\mu,\tau}M_{l}^2+3
\sum_{q=u,d}M_{q}^2)+4 (\sum_{l=e,\mu,\tau}M_{l}^4\nonumber\\
& +3
\sum_{q=u,d}M_{q}^4))-v^2 (3 g_{_W}^2 (M_{H}^2+2
M_{W}^2+M_{Z}^2)+2 \lambda_{1} \nonumber\\
& (3 M_{H}^2+2 M_{W}^2+M_{Z}^2)+2 \lambda_{3} M_{h}^2))\Big]\,\mathcal{C}_{A\phi},
\end{alignat}
and 
\begin{alignat}{4}
\delta\mathcal{C}_{\tilde{A}\phi} & =  \frac{4 g_{_Y}g_{_W}^3M_{H}^2M_{_W}^2+g_{_Y}^3g_{_W}^3M_{H}^2v^2}{64(g_{_Y}^2+g_{W}^2)M_{H}^2M_{Z}^{2}\pi^2}\,\mathcal{C}_{A\tilde{Z}\phi}\nonumber \\ & +\frac{\lambda_{3}\,(g_{_Y}^2\,\mathcal{C}_{\tilde{W}\mathcal{S}}+g_w^2\,\mathcal{C}_{\tilde{B}\mathcal{S}})}{32(g_{_Y}^2+g_{_W}^2)\pi^2}\nonumber\\
& 
+\frac{1}{64 \pi ^2 M_{H}^2 v^2
	(g_{_Y}^2+g_{_W}^2)}\Big[2 g_{_Y}^2 \Big(v^2 (-(g_{_W}^2 \nonumber\\
&  (3(M_{W}^2+M_{Z}^2)-14 M_{H}^2)+\lambda_{1} (3 M_{H}^2+2
M_{W}^2\nonumber\\
& +M_{Z}^2)+\lambda_{3} M_{h}^2))+2 M_{H}^2 (\sum_{l=e,\mu,\tau}M_{l}^2 +3
\sum_{q=u,d}M_{q}^2)\nonumber\\
& +8 (\sum_{l=e,\mu,\tau}M_{l}^4+3
\sum_{q=u,d}M_{q}^4)\Big)-g_{_Y}^4 v^2 (M_{H}^2+3
M_{Z}^2)\nonumber\\
& +g_{_W}^2 (4 (M_{H}^2 (\sum_{l=e,\mu,\tau}M_{l}^2+3
\sum_{q=u,d}M_{q}^2)+4 (\sum_{l=e,\mu,\tau}M_{l}^4\nonumber\\
& +3
\sum_{q=u,d}M_{q}^4))-v^2 (3 g_{_W}^2 (M_{H}^2+2
M_{W}^2+M_{Z}^2)+2 \lambda_{1} \nonumber\\
& (3 M_{H}^2+2 M_{W}^2+M_{Z}^2)+2 \lambda_{3} M_{h}^2))\Big]\,\mathcal{C}_{\tilde{A}\phi}.
\end{alignat}

\bibliography{references} 
%\bibliography{paper.bbl} 

\end{document}